\documentclass[11pt,a4paper]{article}
\pdfoutput=1

\usepackage{jheppub}

\usepackage{amsmath, amssymb, amsthm} 
\usepackage{mathtools}
\usepackage{thmtools}
\usepackage{bm}
\usepackage{dsfont}
\usepackage{braket}
\usepackage{graphicx}

\numberwithin{equation}{section}

\usepackage{enumitem}
\setlist[itemize]{noitemsep}
\setlist[description]{noitemsep}

\usepackage[dvipsnames]{xcolor}
\usepackage{tikz}
\usepackage{pgfplots}
\usetikzlibrary{intersections,backgrounds}
\usepgfplotslibrary{fillbetween}
\pgfplotsset{compat = newest}

\usepackage[numbers, sort&compress]{natbib}
\usepackage{hyperref}
\hypersetup{colorlinks=true, linktoc=page, linkcolor=purple, citecolor=blue}

\newcommand{\negphantom}[1]{
    \ifmmode\settowidth{\dimen0}{$#1$}
    \else\settowidth{\dimen0}{#1}
    \fi
    \hspace*{-\dimen0}}
    
\makeatletter
\newcommand{\mask}[2]{{\mathpalette\mask@{{#1}{#2}}}}
\newcommand{\mask@}[2]{\mask@@{#1}#2}
\newcommand{\mask@@}[3]{%
  \settowidth{\dimen@}{$\m@th#1#2$}%
  \makebox[\dimen@]{$\m@th#1#3$}%
}
\makeatother
    
\newcommand{\sm}{\smallskip}

\usepackage{tikz}
\usetikzlibrary{decorations.pathmorphing,decorations.shapes}
\usetikzlibrary{decorations.pathreplacing,decorations.markings}
\usetikzlibrary{backgrounds}
\usetikzlibrary{positioning}
\usetikzlibrary{arrows}
\usetikzlibrary{shapes,shapes.geometric,shapes.misc}
\usepackage{pgfplots, pgfplotstable}
\pgfplotsset{width=10cm,compat=1.9}

\tikzstyle{tikzfig}=[baseline=-0.25em,scale=0.5]
\pgfkeys{/tikz/tikzit fill/.initial=0}
\pgfkeys{/tikz/tikzit draw/.initial=0}
\pgfkeys{/tikz/tikzit shape/.initial=0}
\pgfkeys{/tikz/tikzit category/.initial=0}

\pgfdeclarelayer{edgelayer}
\pgfdeclarelayer{nodelayer}
\pgfsetlayers{background,edgelayer,nodelayer,main}

\tikzstyle{none}=[inner sep=0mm]

\newcommand{\tikzfig}[1]{%
{\tikzstyle{every picture}=[tikzfig]
\IfFileExists{#1.tikz}
  {\input{#1.tikz}}
  {%
    \IfFileExists{./figures/#1.tikz}
      {\input{./figures/#1.tikz}}
      {\tikz[baseline=-0.5em]{\node[draw=red,font=\color{red},fill=red!10!white] {\textit{#1}};}}%
  }}%
}

\tikzset{->-/.style={decoration={
  markings,
  mark=at position #1 with {\arrow{>}}},postaction={decorate}}}

\tikzset{-<-/.style={decoration={
  markings,
  mark=at position #1 with {\arrow{<}}},postaction={decorate}}}

\tikzstyle{every loop}=[]
\makeatletter
\gdef\@fpheader{}
\makeatother

\begin{document}

\title{Constraining all possible Korteweg–de Vries type hierarchies}

\date{\today}

\author[a]{Lukas W. Lindwasser}
\affiliation[a]{
    Department of Physics\\
    National Taiwan University\\
    Taipei 10617, Taiwan}
\emailAdd{llindwasser@ntu.edu.tw}


\abstract{The Lie algebra of symmetries generated by the left-moving current $j=\partial_-\phi$ in the $2d$ single scalar conformal field theory is infinite dimensional, exhibiting mutually commuting subalgebras. The infinite dimensional mutually commuting subalgebras define integrable deformations of the $2d$ single scalar conformal field theory which preserve the Poisson bracket structure. We study these mutually commuting subalgebras, finding general properties that the generators of such a subalgebra must satisfy. Along the way, we derive constraints on integrable equations of the Korteweg–de Vries type. We also confirm that the recently found $[j]=0,-1,-2$ mutually commuting subalgebras are infinite dimensional.}


\maketitle

 \newpage

\section{Introduction}
\label{sec:intro}
The Korteweg–de Vries equation has a long and interesting history, first written down in 1871 by Boussinesq \cite{Boussinesq} to approximate the behavior of solitary water waves after observations made by Russell \cite{russell1837experimental}, and then further studied by Korteweg and de Vries \cite{Korteweg01051895}. The equation describes the elevation $h-u(x,t)$ ($-u$ small compared to $h$) of the surface of water in a rectangular channel at a horizontal position $x$ and time $t$. In a particular convention of units and parameters, the Korteweg–de Vries equation is
\begin{align}
\label{eq:KdVeqn}
    u_t-6uu_x+u_{xxx}=0
\end{align}
where the subscripts indicate derivatives. Despite it being a nonlinear partial differential equation, exact solutions were found early on, describing a single solitary wave
\begin{align}
\label{eq:solitary}
    u(x,t)=-\frac{c}{2}\text{sech}^2\Big(\frac{\sqrt{c}}{2}(x-ct)\Big)
\end{align}
as well as periodic solutions describing a train of solitary waves \cite{Korteweg01051895}. It was not until 1965 that separated solitary waves (\ref{eq:solitary}) with different velocities were numerically found to appear unaffected by each other after collision \cite{PhysRevLett.15.240}, a property usually associated with linear differential equations. 

\sm

Motivated to understand this peculiar property, the general exact solution to the Korteweg–de Vries equation was constructed via the inverse scattering transform \cite{Gardner:1967wc}, and the equation was subsequently found to exhibit infinitely many commuting charges in a series of papers  \cite{Kruskal1,Kruskal2,10.1063/1.1664873,10.1063/1.1665772,10.1063/1.1665232}, including \cite{Laxorig} where the concept of Lax pairs were introduced. Because of its infinitely many commuting charges, equations of this type are known as integrable equations, harkening back to the notion of Liouville integrability of one dimensional Hamiltonian systems \cite{Liouville1855}.

\sm

Since then many integrable equations, or integrable field theories, have been found in $d=2$ spacetime dimensions, exhibiting particle-like ``soliton" solutions. These solvable theories have offered a useful non-perturbative laboratory to understand phenomenologically relevant concepts, e.g., \cite{PhysRevD.11.3424,PhysRevD.11.2088,ZAMOLODCHIKOV1979253,Dubovsky:2015zey,Copetti:2024rqj}. Beyond this, $2d$ integrable models appear commonly within string theory \cite{DHoker:1982wmk,Metsaev:1998it,PhysRevD.69.046002,Beisert_2011} and, more broadly, in gauge theory \cite{Ablowitz1993,Costello_2018,Costello:2018gyb,Costello:2019tri,Gutperle:2014aja,Beccaria:2015iwa}. There are also deep connections between $2d$ integrable models and $2d$ conformal field theory \cite{Sasaki:1987mm,Eguchi:1989hs,ZAMOLODCHIKOV1989641,Bazhanov:1994ft,Smirnov:2016lqw}. 

\sm

Given the historical significance the Korteweg–de Vries equation has on this field, and the wealth of techniques that were subsequently developed to solve nonlinear partial differential equations, it is important to understand what other integrable equations are of the generic form
\begin{align}
\label{eq:inteqn}
    &u_t - \left(V(\{u,u_x,u_{xx},\dots\},x)\right)_x=0, && V=\frac{\delta H}{\delta u(x,t)}
\end{align}
where we will call $V$ the potential. In the case of the Korteweg–de Vries equation (\ref{eq:KdVeqn}), $V_{\text{KdV}}=3u^2-u_{xx}$. Equations of this type have a natural Poisson structure
\begin{align}
\label{eq:Poisson}
    \{F,G\}\equiv\int dx\frac{\delta F}{\delta u(x,t)}\frac{d}{dx}\frac{\delta G}{\delta u(x,t)}
\end{align}
where $\frac{\delta}{\delta u(x,t)}$ is a functional derivative with respect to $u(x,t)$. The functional $H$ in (\ref{eq:inteqn}) is the Hamiltonian of this system with respect to this Poisson bracket. We are therefore searching for potentials $V$ that result in an integrable equation, having infinitely mutually commuting charges in the sense of (\ref{eq:Poisson}).

\sm

Any potential $V=V(u)$ that does not depend on derivatives of $u$ or explicitly on $x$ results in an integrable equation, because any other function $U(u)$ is an integrating factor, implying a conserved current for each such $U(u)$,
\begin{align}
   &0= U(u)\big(u_t-(V(u))_x\big)\\
    &0=U(u)u_t-U(u)V'(u)u_x\\
    &0=\frac{d}{dt}\Big(\int du U(u)\Big)-\frac{d}{dx}\Big(\int du U(u)V'(u)\Big)
\end{align}
and the charges implied by each current can be shown to all Poisson commute with each other.

\sm

We are mainly concerned with potentials $V=V(\{u,u_x,u_{xx},\dots\})$ that are allowed to depend on derivatives of $u$. In particular, we are concerned with potentials that are derivable from an action principle in a similar fashion to the Korteweg–de Vries equation, that are order $2n$ in $x$ derivatives of $u$. These potentials are always linear in $u_{2n}$, where $u_i$ is the $i$th $x$ derivative of $u$, which we write generically as
\begin{align}
\label{eq:quasilinearV}
    V\left(\{u,u_x,\dots,u_{2n}\}\right)=A(u,u_x,\dots,u_n)u_{2n}+B(u,u_x,\dots,u_{2n-1})
\end{align}
where $A$ is an order $n$ functional of $u$, and $B$ is an order $2n-1$ functional of $u$, and is a polynomial in $u_{n+1},\dots,u_{2n-1}$. As an example, by defining $u=\phi_x$, the Korteweg–de Vries equation (\ref{eq:KdVeqn}) can be obtained from the following action by varying with respect to $\phi$
\begin{align}
    S_{\text{KdV}}=\int dxdt\left(\phi_t\phi_x-\phi_{xx}^2-2\phi_x^3\right)
\end{align}
For this reason we refer to the potentials (\ref{eq:quasilinearV}) as Korteweg–de Vries type potentials.

\sm

It was clear early on that not any Korteweg–de Vries type potential results in an integrable equation. For instance, it was observed that an equation with a potential of the form $V=\alpha u^{n}-u_{xx}$ is integrable only when $n=2,3$ \cite{Kruskal1}, with $n=3$ corresponding to the modified Korteweg–de Vries equation. 

\sm

Demonstrating integrability for a generic Korteweg–de Vries type potential $V$ is generally difficult. The few examples so far mentioned exhibit nontrivial Lax pairs, whose existence prove integrability. Lax pairs, however, do not necessarily exist whenever an equation is integrable, and so other techniques for proving integrability are needed in this setting. Instead, the more general notion of a formal recursion operator $\Lambda$, which is a pseudo-differential operator that satisfies an equation $\Lambda_t=[V_*,\Lambda]$, where $V_*$ is an operator associated with the potential, necessarily exists when $V$ results in an integrable equation. A psuedo-differential operator is an operator that is formally expanded as a sum of terms with derivatives up to some order $r$, including possible negative orders
\begin{align}
    \Lambda = \sum_{n=-\infty}^r\Lambda_n(\{u,u_x,u_{xx},\dots\},x)\partial_x^n
\end{align}
As an example, the Korteweg–de Vries equation (\ref{eq:KdVeqn}) has the formal recursion operator
\begin{align}
    \Lambda_{\text{KdV}}=\partial_x^2-4u-2u_x\partial_x^{-1}
\end{align}
whereas the modified Korteweg–de Vries equation, with a potential $V_{\text{mKdV}}=2u^3-u_{xx}$, has the formal recursion operator
\begin{align}
    \Lambda_{\text{mKdV}}=\partial_x^2-4u^2-4u_x\partial_x^{-1}(u\,\cdot\,)
\end{align}
Once one has a nontrivial formal recursion operator $\Lambda$, the infinitely many charges are obtained by taking (not necessarily integer) powers $\Lambda^k$ and extracting the coefficient in front of $\partial_x^{-1}$. For a generic Korteweg–de Vries type potential $V$, it is difficult to solve for $\Lambda$. An old approach known as the ``symmetry approach" to classifying integrable equations of the type (\ref{eq:inteqn}) uses the existence of $\Lambda$, solving explicitly for enough $\Lambda_n$, providing constraints on $V$ that reduce the list of possible integrable equations of a given order to a finite number up to ``almost invertible" transformations \cite{SISvinolupov_1992}. By now there is a complete classification of potentials of order 2, and a classification of potentials of order 4 with constant separant\footnote{A Korteweg–de Vries type potential of order $2n$ with constant separant takes the form $V(\{u,u_x,\dots,u_{2n}\})=B(\{u,u_x,\dots,u_{2n-1}\})-u_{2n}$.} up to almost invertible transformations \cite{SISvinolupov_1992,Svinolupov1982,Sokolov1985,Mikhailov1991,Mikhailov2009,meshkov2013integrable}, summarized relatively recently in \cite{Heredero:2019arc} with more references therein.  Integrability of a given potential $V$ can be proven after showing that the resulting equation is equivalent to one of the integrable equations on these lists. A complete classification of integrable equations of the type (\ref{eq:inteqn}) is however still needed. 

\sm

Although there are constraints on potentials of this type, there are many integrable equations of this form already known. Indeed, associated with the Korteweg–de Vries equation is an \textit{infinite hierarchy} of integrable equations, known as higher Korteweg–de Vries equations. The first few have potentials
\begin{align}
    V_{\text{KdV},1}&=3u^2-u_{xx}\\
    V_{\text{KdV},2}&=10u^3-5u_x^2-10uu_{xx}+u_{xxxx}\\
    V_{\text{KdV},3}&=35u^4-70uu_x^2-70u^2u_{xx}+21u_{xx}^2+28u_xu_{xxx}+14uu_{xxxx}-u_{xxxxxx}\\
    &\;\;\vdots\nonumber
\intertext{and so on. All of the integrable equations in a hierarchy share the same conserved charges. The potential for the modified Korteweg–de Vries equation $V_{\text{mKdV}}=2u^3-u_{xx}$ also has an infinite hierarchy of integrable equations, with the first few potentials}
    V_{\text{mKdV},1}&=2u^3-u_{xx}\\
    V_{\text{mKdV},2}&=6u^5-10uu_x^2-10u^2u_{xx}+u_{xxxx}\\
    V_{\text{mKdV},3}&=20u^7-140u^3u_x^2-70u^4u_{xx}+70u_x^2u_{xx}+42uu_{xx}^2+56uu_xu_{xxx}\nonumber\\
    &\;\;\;\;+14u^2u_{xxxx}-u_{xxxxxx}\\
    &\;\;\vdots\nonumber
\end{align}
and so on. In fact, as we will review in \autoref{ssec:KdVhierarchies}, \textit{any} Korteweg–de Vries type potential that results in an integrable equation has an infinite hierarchy of integrable equations associated with it. To simplify the discussion, we may then treat each hierarchy as a single set of integrable equations, identifying them with their potential which has the lowest maximum nonzero order of derivatives of $u$ in the hierarchy.

\sm

The purpose of this paper is to further develop an alternative approach to classifying $2d$ integrable models introduced in \cite{Lindwasser:2024qyh}, which can find within this approach necessary conditions on the lowest maximum nonzero order potential $V$ in each hierarchy, narrowing in on the parameter space of integrable equations of the type (\ref{eq:quasilinearV}) and also other types of integrable equations. This approach is complimentary to the older ``symmetry approach", and we will see how we can arrive at new constraints on Korteweg–de Vries type potentials of \textit{any} order \textit{without} constant separant.

To do this, we reformulate the problem as a Lie algebra problem as in \cite{Lindwasser:2024qyh}. As we explain in more detail in \autoref{sec:Preliminaries}, equations of the type (\ref{eq:quasilinearV}) can be obtained from particular deformations of the $2d$ single scalar conformal field theory which preserve an infinite number of mutually commuting symmetries
\begin{align}
\label{eq:scalarlambdaaction}
    S=\int d^2x\left(\partial_+\phi\partial_-\phi - \lambda \,\ell(\{j_{n}\},x^-)\right)
\end{align}
where $\ell(\{j_{n}\},x^-)$ is a functional of $j=\partial_-\phi$ and its $x^-$ derivatives up to order $n$. The equation of motion of this action takes the form (\ref{eq:quasilinearV}) after making the identifications $x^+\to t$, $x^-\to x$, $j\to u$, and $\frac{\lambda}{2}\mathcal{E}(\ell)\to V$, where $\mathcal{E}(\ell)$ is the Euler–Lagrange equation of $\ell(\{j_n\},x^-)$, calculated as if it were a $1d$ Lagrangian of order $n$.

\sm

The Lie algebra of symmetries of the action (\ref{eq:scalarlambdaaction}) when $\lambda=0$ is infinite dimensional, with infinite dimensional mutually commuting subalgebras. We will see how each such subalgebra defines an integrable hierarchy, and so the problem at hand becomes finding all infinite dimensional mutually commuting subalgebras. In a similar fashion to the symmetry approach of classifying integrable equations \cite{Heredero:2019arc}, in the absence of simple structures that prove infinite dimensionality like the existence of a Lax pair, we will only be able to find necessary conditions on infinite dimensional mutually commuting subalgebras. In \autoref{sec:Summary}, we will summarize the necessary conditions on these subalgebras found here and in \cite{Lindwasser:2024qyh}. 

\sm

The infinite dimensional mutually commuting subalgebras of the $2d$ single scalar conformal field theory generated by infinitesimal transformations $\delta_i$ such that $[\delta_i,\delta_j]=0$ for all $i,j$ may admit integrable deformations, by which we mean deformations to the action $S\to S+S_{\text{deform}}$ satisfying $\delta_{i}S_{\text{deform}}=0$  for all $i$, which are not of the form (\ref{eq:scalarlambdaaction}), like the deformations defining the Liouville, sine-Gordon, and Bullough–Dodd (or Tzitz\'{e}ica) models. In this way, studying these subalgebras is more general than classifying integrable equations of the type (\ref{eq:quasilinearV}). One may think of this paper more simply as studying properties of the Lie algebra of symmetries present in the $2d$ single scalar conformal field theory. Constraints on integrable deformations are then obtained from this analysis.

\sm

Formulating this problem through the Lie algebra of symmetries present in the $2d$ single scalar conformal field theory makes it clear that the same exercise can be performed on a much broader class of theories. In particular, any $2d$ action with a left(right)-moving current, like a collection of scalars, a collection of left- (right-) moving fermions, Wess–Zumino–Witten models, or principal chiral models, has an infinite dimensional Lie algebra of symmetries. In the same way as the $2d$ single scalar conformal field theory, the infinite dimensional mutually commuting subalgebras of all of these theories define integrable hierarchies. By studying the Lie algebras associated with all such theories, one can find a large class of integrable models. We study here the case of the single scalar.
\subsection{Conventions}
\label{sec:conv}
We will mostly work in this paper at the action level, using the variables $\phi$ and $j=\partial_-\phi$. For brevity, we will often use the shorthand $j_k\equiv \partial_-^kj$. Working in Minkowski space, we use light-cone coordinates with the convention
\begin{align}
    x^{\pm}=\frac{1}{\sqrt{2}}(x^0\pm x^1),\qquad \partial_{\pm}=\frac{1}{\sqrt{2}}(\partial_0\pm \partial_1)
\end{align}
We will also on occasion go back and forth between these conventions, and the standard conventions surrounding the Korteweg–de Vries equation
\begin{align}
\label{eq:KdVconv}
    &x^+\longleftrightarrow t && x^-\longleftrightarrow x && j\longleftrightarrow u
\end{align}
\subsection{Outline}
\label{sec:outline}
In \autoref{sec:Preliminaries}, we will review the Lie algebra of symmetries present in the classical $2d$ single scalar conformal field theory in terms of generators defined on the space of $1d$ Lagrangians $\ell(\{j(x^-),\partial_-j(x^-),\dots\},x^-)$, and show how its infinite dimensional mutually commuting subalgebras define integrable hierarchies of the type (\ref{eq:quasilinearV}). We will also confirm in \autoref{ssec:mutcommutsub} that the mutually commuting subalgebras corresponding to $[j]=0,-1,-2$ found in \cite{Lindwasser:2024qyh} are infinite dimensional. Then in \autoref{sec:Summary}, we summarize the necessary conditions on Korteweg–de Vries type hierarchies found here and in \cite{Lindwasser:2024qyh}. In particular we describe the main result of this paper, which is the necessary $j_{n_1}$ dependence of the lowest nonzero order $n_1$ Lagrangian $\ell_{n_1}$ in any infinite dimensional mutually commuting subalgebra when $n_1=1,2,3,4$ and, more generally, when $n_1>4$ depending on certain technical details. In \autoref{sec:argument}, we provide a proof for this. Finally, in \autoref{sec:discussion}, we discuss our results and present some questions for future research.
\section{Preliminaries}
\label{sec:Preliminaries}
In this section, we review the Lie algebra of symmetries present in the classical $2d$ single scalar conformal field theory. We also review how its infinite dimensional mutually commuting subalgebras define integrable hierarchies of the Korteweg–de Vries type, and briefly review the mutually commuting subalgebras that are currently known.

\sm

The $2d$ single scalar conformal field theory has the action
\begin{align}
\label{eq:freeaction}
    S = \int d^2x\partial_+\phi\partial_-\phi
\end{align}
Due to this action's Euler–Lagrange equation $\partial_+\partial_-\phi=0$, this theory has a left-moving current $j(x^-)=\partial_-\phi$ satisfying $\partial_+j(x^-)=0$ (and a right-moving current $\partial_+\phi$).

\sm

The existence of a left-moving current $j(x^-)$ is very constraining on the dynamics of the theory, because it implies the existence of infinitely many left-moving currents $\ell(\{j(x^-),\partial_-j(x^-),\dots\},x^-)$ that are arbitrary \textit{functionals} of $j(x^-)$. It is useful to assume that these functionals have a maximum finite order $k$ derivative of $j(x^-)$ which the functional depends on. In this case, we will use the shorthand $\ell(\{j_k(x^-)\},x^-)$ or more often simply as $\ell_k$, to indicate such a functional, where $j_k(x^-)\equiv \partial_-^kj(x^-)$. The Lie algebra is generated by the charges associated with these currents
\begin{align}
    Q_{\ell} = \int dx^-\ell(\{j_k(x^-)\},x^-)
\end{align}
Assuming that $j(x^-)$ satisfies vanishing (at infinity) or periodic boundary conditions, unless it has a term that is independent of $j(x^-)$, any functional that is a total derivative $\ell = \partial_- f(\{j_k(x^-)\},x^-)$ results in a vanishing charge. The set of functionals that generate the algebra then is the set of all functionals modulo total derivatives or, in other words, the set of all $1d$ Lagrangians of $j(x^-)$. The Lie algebra then acts on the vector space of $1d$ Lagrangians.

\sm

We will say that two $1d$ Lagrangians are equivalent $\ell \sim\ell'$ if and only if they generate the same Euler–Lagrange equation, i.e., they differ by a total derivative $\ell = \ell' + \partial_-f$. A $1d$ Lagrangian $\ell(\{j_k(x^-)\},x^-)$ that depends on $j(x^-)$ and its derivatives up to order $k$ may be equivalent to a Lagrangian $\ell'(\{j_{k'}(x^-)\},x^-)$ with $k'<k$. We define the \textit{order of a Lagrangian} to be the minimum value of $k$ ($k$ being the maximum order derivative of $j(x^-)$ that $\ell$ depends on) that can be obtained by adding total derivatives. As a simple example, $\ell(\{j_6(x^-)\},x^-)=j(x^-) j_6(x^-)$ is order 3, because $jj_6\sim-j_1j_5\sim j_2j_4\sim-j_3j_3$. We will assume without loss of generality that $\ell(\{j_k(x^-)\},x^-)$ has been reduced such that $k$ represents the order of $\ell$.
\subsection{Lie algebra}
Each $1d$ Lagrangian $\ell(\{j_i(x^-)\},x^-)$ generates an infinitesimal symmetry transformation of the action (\ref{eq:freeaction})
\begin{align}
    \label{eq:scalarsym}
&\phi\to\phi+\epsilon\delta_{\ell}\phi\\
\label{eq:scalarsym1}
    &\delta_{\ell}\phi = \frac{1}{2}\mathcal{E}(\ell) = \frac{1}{2}\sum_{k=0}^{\infty}(-1)^k\partial_-^k\frac{\partial}{\partial j_k(x^-)}\ell(\{j_i(x^-)\},x^-)
\end{align}
where $\mathcal{E}$ is a linear operator called the Euler operator, for which $\mathcal{E}(\ell)$ gives the Euler–Lagrange equation for $j(x^-)$, as derived from the $1d$ Lagrangian $\ell(\{j_i(x^-)\},x^-)$, and $\epsilon$ is some infinitesimal parameter. Since the Lagrangian is of finite order, the infinite sum in (\ref{eq:scalarsym1}) truncates at $k=i$. The Euler operator has the important property that $\mathcal{E}(\ell)=0$ for any value of $j(x^-)$ and its derivatives if and only if $\ell=\partial_-f\sim 0$ is a total derivative \cite{olver1993applications}. We will use this property repeatedly throughout this paper.

\sm

The infinitesimal symmetry transformations (\ref{eq:scalarsym1}) associated with two Lagrangians $\ell$ and $\ell'$ do not, in general, commute. By direct computation, the commutator $[\delta_{\ell},\delta_{\ell'}]\phi$ is 
\begin{align}
\label{eq:scalarliedirect}
    [\delta_{\ell},\delta_{\ell'}]\phi = \frac{1}{4}\sum_{k=0}^{\infty}\left(\partial_-^{k+1}\mathcal{E}(\ell)\frac{\partial}{\partial j_k(x^-)}\mathcal{E}(\ell')-\partial_-^{k+1}\mathcal{E}(\ell')\frac{\partial}{\partial j_k(x^-)}\mathcal{E}(\ell)\right)
\end{align}
 We may write the right-hand side of this equation in the form $\frac{1}{2}\mathcal{E}(\ell'')$, where $\ell''$ is a new Lagrangian
\begin{align}
\label{eq:scalarlie}
[\delta_{\ell},\delta_{\ell'}]\phi=\delta_{\ell''}\phi=\frac{1}{2}\mathcal{E}(\ell''),\qquad \ell''=\frac{1}{4}\left(\partial_-\mathcal{E}(\ell)\mathcal{E}(\ell')-\partial_-\mathcal{E}(\ell')\mathcal{E}(\ell)\right)
\end{align}
where $\ell''$ is written in a manifestly anti-symmetric in $(\ell\leftrightarrow \ell ')$ way; although, because Lagrangians are defined modulo total derivatives, we may write $\ell ''=\frac{1}{2}\partial_-\mathcal{E}(\ell)\mathcal{E}(\ell')$. The commutator (\ref{eq:scalarlie}) defines a Lie algebra on the vector space of $1d$ Lagrangians. If the commutator Lagrangian $\ell''=\partial_-f\sim 0$ is a total derivative, then $\delta_{\ell''}\phi=\frac{1}{2}\mathcal{E}(\ell'')=0$ is a trivial transformation, and we say that $\ell$ and $\ell'$ commute. 
\subsection{Korteweg–de Vries type hierarchies}
\label{ssec:KdVhierarchies}
Suppose there is an infinite dimensional mutually commuting subalgebra of (\ref{eq:scalarlie}) with generators $\ell_{n_i}(\{j_{n_i}(x^-)\},x^-)$ of order $n_i$ for $i=1,2,3,\dots,\infty$ satisfying
\begin{align}
\label{eq:KdVcommute}
    \frac{1}{4}\left(\partial_-\mathcal{E}(\ell_{n_i})\mathcal{E}(\ell_{n_j})-\partial_-\mathcal{E}(\ell_{n_j})\mathcal{E}(\ell_{n_i})\right)\sim 0
\end{align}
for any $i,j$. Such a subalgebra implies the existence of an infinite hierarchy of integrable models, each model defined by deforming (\ref{eq:freeaction}) with $\ell_{n_i}(\{j_{n_i}(x^-)\},x^-)$,
\begin{align}
\label{eq:KdVtypeaction}
    S_{\ell_{n_i}}=\int d^2x\left(\partial_+\phi\partial_-\phi - \lambda \,\ell_{n_i}(\{j_{n_i}\},x^-)\right)
\end{align}
This action is integrable for any $n_i$, because it is invariant under the infinitesimal transformation $\delta_{\ell_{n_j}}\phi=\frac{1}{2}\mathcal{E}(\ell_{n_j})$ for all $i,j$ and finite $\lambda$. Indeed, the kinetic term is invariant under any $1d$ Lagrangian transformation, while $\delta_{\ell_{n_j}}\ell_{n_i}\sim 0$ for all $i,j$ because
\begin{align}
    \delta_{\ell_{n_j}}\ell_{n_i}(\{j_{n_i}\},x^-)&=\frac{1}{2}\sum_{k=0}^{\infty}\partial_-^{k+1}(\mathcal{E}(\ell_{n_j}))\frac{\partial}{\partial j_k}\ell_{n_i}(\{j_{n_i}\},x^-) \nonumber \\
    &\sim \frac{1}{2}\sum_{k=0}^{\infty}(-1)^k\partial_-\mathcal{E}(\ell_{n_j})\partial_-^{k}\frac{\partial}{\partial j_k}\ell_{n_i}(\{j_{n_i}\},x^-) \nonumber \\
    &=\frac{1}{2}\partial_-\mathcal{E}(\ell_{n_j})\mathcal{E}(\ell_{n_i})\sim 0
\end{align}

\sm

The equations of motion for the action (\ref{eq:KdVtypeaction}) can be written entirely in terms of $j(x^+,x^-)=\partial_-\phi(x^+,x^-)$,
\begin{align}
\label{eq:KdVtypeeq}
    \partial_+j-\frac{\lambda}{2}\partial_-\mathcal{E}(\ell_{n_i})=0
\end{align}
In this way, each $\ell_{n_i}$ defines a potential $V=\frac{\lambda}{2}\mathcal{E}(\ell_{n_i})$ for an integrable equation of the form (\ref{eq:quasilinearV}). Momentarily matching the standard conventions (\ref{eq:KdVconv}), the potential $V$ is order $2n_i$ in $x$ derivatives of $u$ and is linear in $u_{2n_i}$
\begin{align}
\label{eq:quasilinearpotential}
    V(\{u,u_x,\dots,u_{2n_i}\})=A(u,u_x,\dots,u_{n_i})u_{2n_i}+B(u,u_x,\dots,u_{2n_i-1})
\end{align}

The classic example of an integrable hierarchy of this type is the Korteweg–de Vries equation and its infinite set of higher Korteweg–de Vries equations. This hierarchy is associated with an infinite dimensional mutually commuting subalgebra, first proven to be infinite dimensional in \cite{Kruskal2}, the first six generators $\ell_{n_i}$ of which we list here
\begin{align}
\label{eq:KdVell1}
  &\ell_1:  &&j_1^2+2j^3,\\
   &\ell_2:  &&j_2^2+10jj_1^2+5j^4,\\
  &\ell_3:  &&j_3^2+14jj_2^2+70j^2j_1^2+14j^5,\\
  \label{eq:KdVell4}
  &\ell_4:  &&j_4^2+18jj_3^2-20j_2^3+126j^2j_2^2-35j_1^4+420j^3j_1^2+42j^6, \\
  &\ell_5:  &&j_5^2 +22jj_4^2-110j_2j_3^2+198j^2j_3^2-440jj_2^3-462j_1^2j_2^2\nonumber \\
    \label{eq:KdVell5}
  &  && +924j^3j_2^2-770jj_1^4 +2310j^4j_1^2+132j^7, \\
  \label{eq:KdVell6}
  &\ell_6: &&j_6^2 + 26 jj_5^2 - 182j_2j_4^2+286j^2j_4^2 - 2860jj_2j_3^2 - 858j_1^2j_3^2 \nonumber \\
  & && +1716j^3j_3^2 + 1001j_2^4 - 5720j^2j_2^3 - 12012jj_1^2j_2^2 + 6006j^4j_2^2 \nonumber \\
  & && - 10010j^2j_1^4 +12012j^5j_1^2 + 429j^8
\end{align}
The Korteweg–de Vries equation is obtained from (\ref{eq:KdVtypeeq}) with $\ell_1=j_1^2+2j^3$. Setting $\lambda=1$, and matching the standard conventions (\ref{eq:KdVconv}), the Korteweg–de Vries equation is
\begin{align}
\label{eq:KdVeq}
    u_t-6uu_x+u_{xxx}=0
\end{align}
where the subscripts indicate derivatives.

\sm

We will call any integrable hierarchy associated with an infinite dimensional mutually commuting subalgebra of (\ref{eq:scalarlie}) a Korteweg–de Vries type hierarchy.
\subsection{Infinite dimensional mutually commuting subalgebras}
\label{ssec:mutcommutsub}
There are many mutually commuting subalgebras of (\ref{eq:scalarlie}) known so far. We have already mentioned one infinite dimensional subalgebra above, with the first six generators (\ref{eq:KdVell1})–(\ref{eq:KdVell6}). Having more explicit examples of infinite dimensional mutually commuting subalgebras at our disposal is useful as a guide for understanding their general structure. In this subsection, we review the other infinite dimensional mutually commuting subalgebras, listing only the first three generators of each subalgebra for brevity. In particular, we will confirm in this subsection that the new mutually commuting subalgebras found in \cite{Lindwasser:2024qyh} are also infinite dimensional.

\sm

These subalgebras were found assuming the generators $\ell_i$ have definite scaling dimension $[\ell_i]$, assigning $x^-$ with a scaling dimension $[x^-]=-1$ and the current $j(x^-)$ with some scaling dimension $[j]$, which is not necessarily the canonical scaling dimension 1. In particular, these subalgebras were found with the choices $[j]=0,\pm 1,\pm 2$. The subalgebra associated with the Korteweg–de Vries model (\ref{eq:KdVell1})–(\ref{eq:KdVell6}) has definite scaling dimension generators when $[j]=2$.

\sm

When $[j]=1$, there are two distinct definite scaling dimension mutually commuting subalgebras (unique up to an affine transformation $j\to aj+b$ with $a,b\in\mathbb{C}$). The first subalgebra corresponds to the symmetries of Liouville, sine-Gordon, and modified Korteweg–de Vries models, also proven to be infinite dimensional in \cite{Kruskal2}
\begin{align}
\label{eq:mKdVell1}
   &\ell_1: &&j_1^2+j^4, \\
   &\ell_2: &&j_2^2+10j^2j_1^2+2j^6,\\
\label{eq:mKdVell3}
   &\ell_3: &&j_3^2+14j^2j_2^2-7j_1^4+70j^4j_1^2+5j^8\\
\intertext{This subalgebra includes Lagrangians $\ell_{n_i}$ of all orders $n_i=i$ where $i=1,2,3,\dots$. With the convention (\ref{eq:KdVconv}), the lowest order generator $\ell_1$ results in the modified Korteweg–de Vries equation}
    & && \qquad\qquad\qquad\qquad u_t-6u^2u_x+u_{xxx}=0
\intertext{The second subalgebra with $[j]=1$ corresponds to the symmetries of the Tzitz\'{e}ica model, originally discovered in \cite{zbMATH02644938} and proven to be infinite dimensional in \cite{Zhiber:1979am,Mikhailov1981}}
\label{eq:BDell2}
  &\ell_2:  &&j_2^2-5j_1^3+45j^2j_1^2+27j^6,\\
  &\ell_3:  &&j_3^2-21j_1j_2^2+63j^2j_2^2-21j_1^4-126j^2j_1^3+1134j^4j_1^2+243j^8, \\
\label{eq:BDell5}
  &\ell_5:  &&j_5^2-33j_1j_4^2+44j_3^3+99j^2j_4^2-990jj_2j_3^2-594j_1^2j_3^2+561j_2^4 +8118j_1^3j_2^2\nonumber \\
  &  && +3168jj_1j_2^3-1782j^2j_1j_3^2+\frac{24156}{5}j_1^6 - 58806j^2j_1^2j_2^2-15444j^3j_2^3 \nonumber \\
  &  &&+3564j^4j_3^2 + 44550j^2j_1^5-26730j^4j_1j_2^2 + 58806j^6j_2^2-267300j^4j_1^4 \nonumber \\
  &  && - 53460j^6j_1^3 + 481140j^8j_1^2 +26244j^{12}
\end{align}
This subalgebra includes Lagrangians $\ell_{n_i}$ of all orders such that $n_i\equiv0,2\,(\text{mod }3)$. The lowest order generator $\ell_2$ results in the integrable equation
\begin{align}
    u_t - 405u^4u_x+45u_x^3+180uu_xu_{xx}-15u_{xx}^2+45u^2u_{xxx}-15u_xu_{xxx}-u_{xxxxx}=0
\end{align}

\sm

When $[j]=0$, there is one distinct definite scaling dimension mutually commuting subalgebra (again unique up to an affine transformation $j\to aj+b$ with $a,b\in\mathbb{C}$)
\begin{align}
\label{eq:j=0ell1}
    &\ell_1: &&\frac{j_1^2}{(1-j^2)^3}, \\
\label{eq:j=0ell2}
    &\ell_2: &&\frac{j_2^2}{(1-j^2)^5}-\frac{5}{3}\frac{1+8j^2}{(1-j^2)^7}j_1^4,\\
\label{eq:j=0ell3}
    &\ell_3: &&\frac{j_3^2}{(1-j^2)^7}-14\frac{j}{(1-j^2)^8}j_2^3-14\frac{2+13j^2}{(1-j^2)^9}j_1^2j_2^2 +\frac{14}{5}\frac{11+248j^2+416j^4}{(1-j^2)^{11}}j_1^6
\intertext{ This subalgebra appears to follow the same pattern of orders $n_i$ as the subalgebra associated with among other things the Liouville model with generators (\ref{eq:mKdVell1})–(\ref{eq:mKdVell3}). The lowest order generator $\ell_1$ results in the equation}
\label{eq:KdVj=0}
& &&\qquad u_t+3\frac{1+7u^2}{(1-u^2)^5}u_x^3+12\frac{u}{(1-u^2)^4}u_xu_{xx}+\frac{1}{(1-u^2)^3}u_{xxx}=0
\intertext{To show that this equation is integrable, and that it therefore has infinitely many independent symmetries generated by a mutually commuting subalgebra containing the generators (\ref{eq:j=0ell1})–(\ref{eq:j=0ell3}), it is sufficient to show that (\ref{eq:KdVj=0}) is equivalent to one of the integrable equations in the classification schemes found for instance in \cite{Svinolupov1982}. This equation is indeed equivalent after performing potentiation and point transformations to}
\label{eq:intclass1}
& && \hspace{4cm}u_t -\frac{3}{2}\frac{u_xu_{xx}^2}{u_x^2-1} + u_{xxx}=0
\intertext{which belongs to one of the classes of third order integrable equations with constant separant originally found in \cite{Svinolupov1982}. We show this in a series of steps, first by making the point transformation $\tilde{u}=1-u^2$ on (\ref{eq:KdVj=0}), resulting in the equation}
\label{eq:inttran1}
& &&\hspace{3cm}\tilde{u}_t=\partial_x\left(-\frac{\tilde{u}_{xx}}{\tilde{u}^3}+\frac{3}{4}\frac{3\tilde{u}-2}{(\tilde{u}-1)\tilde{u}^4}\tilde{u}_x^2\right)
\intertext{The next step is potentiation, defining $\tilde{u}=\partial_x\hat{u}$, we get the equation}
\label{eq:inttran2}
& && \hspace{3cm}\hat{u}_t=-\frac{\hat{u}_{xxx}}{\hat{u}_x^3}+\frac{3}{4}\frac{3\hat{u}_x-2}{(\hat{u}_x-1)\hat{u}_x^4}\hat{u}_{xx}^2
\intertext{Next we make another point transformation, redefining $x=u$ and $\hat{u}=\hat{x}$}
\label{eq:inttran3}
& && \hspace{3cm}u_t+\frac{3}{4}\frac{1-2u_{\hat{x}}}{(u_{\hat{x}}-1)u_{\hat{x}}}u_{\hat{x}\hat{x}}^2+u_{\hat{x}\hat{x}\hat{x}}=0
\intertext{Relabeling $\hat{x}=x$, and performing the last transformation $u\to \frac{1}{2}(u+x)$, we arrive at (\ref{eq:intclass1}).
\newline
\sm
\hspace{.72cm}When $[j]=-1$, there is one distinct definite scaling dimension mutually commuting subalgebra (unique up to an affine transformation $j_1\to aj_1+b$ with $a,b\in\mathbb{C}$)}
\label{eq:j=-1ell1}
    &\ell_1: &&(1+j_1^2)^{1/2}, \\
\label{eq:j=-1ell2}
    &\ell_2: &&\frac{j_2^2}{(1+j_1^2)^{5/2}},\\
    \label{eq:j=-1ell3}
    &\ell_3: &&\frac{j_3^2}{(1+j_1^2)^{7/2}}+\frac{7}{4}\frac{1-4j_1^2}{(1+j_1^2)^{11/2}}j_2^4
\intertext{This subalgebra appears to follow the same pattern of orders $n_i$ as the subalgebra with generators (\ref{eq:mKdVell1})–(\ref{eq:mKdVell3}) and the subalgebra with generators (\ref{eq:j=0ell1})–(\ref{eq:j=0ell3}). The lowest order generator $\ell_1$ results in the equation}
\label{eq:KdVj=-1}
& &&\hspace{2.5cm} u_t -\frac{3}{2}\frac{u_xu_{xx}^2}{(1+u_x^2)^{5/2}} + \frac{1}{2}\frac{u_{xxx}}{(1+u_x^2)^{3/2}}=0
\intertext{
Interestingly, one can show that this equation is also equivalent after performing potentiation and point transformations to (\ref{eq:intclass1}). This can be shown in a similar fashion to the steps performed in (\ref{eq:inttran1})–(\ref{eq:inttran3}). So although at the level of the Lie algebra, the equations (\ref{eq:KdVj=0}) and (\ref{eq:KdVj=-1}) are associated with distinct subalgebras, they are of the same class of third order integrable equations in the canonical classification scheme defined up to ``almost invertible" transformations \cite{SISvinolupov_1992}. This demonstrates that (\ref{eq:KdVj=-1}) is integrable, and that it has infinitely many independent symmetries generated by a mutually commuting subalgebra containing the generators (\ref{eq:j=-1ell1})–(\ref{eq:j=-1ell3}).
\newline
\sm
\hspace{.72cm}And finally when $[j]=-2$, there is one distinct definite scaling dimension mutually commuting subalgebra (unique up to an affine transformation $j_2\to aj_2+b$ with $a,b\in\mathbb{C}$)} 
\label{eq:j=-2ell2}
    &\ell_2: &&(1+j_2)^{1/3}, \\
\label{eq:j=-2ell3}
    &\ell_3: &&\frac{j_3^2}{(1+j_2)^{7/3}},\\
\label{eq:j=-2ell5}
    &\ell_5: &&\frac{j_5^2}{(1+j_2)^{11/3}} + \frac{44}{9}\frac{j_4^3}{(1+j_2)^{14/3}}-\frac{220}{9}\frac{j_3^2j_4^2}{(1+j_2)^{17/3}} + \frac{6545}{243}\frac{j_3^6}{(1+j_2)^{23/3}}
\end{align}
This subalgebra appears to follow the same pattern of orders $n_i$ as the subalgebra associated with the Tzitz\'{e}ica model with generators (\ref{eq:BDell2})–(\ref{eq:BDell5}). In this case, the lowest order generator $\ell_2$ results in the equation
\begin{align}
    u_t+\frac{40}{81}\frac{u_{xxx}^3}{(1+u_{xx})^{11/3}}-\frac{5}{9}\frac{u_{xxx}u_{xxxx}}{(1+u_{xx})^{8/3}}+\frac{1}{9}\frac{u_{xxxxx}}{(1+u_{xx})^{5/3}}=0
\end{align}
This equation is integrable, because it is equivalent after performing potentiation, point transformations and field redefinitions to 
\begin{align}
    u_t +cu_x+ u_x^5-5u_xu_{xx}^2+5\left(u_{xx}-u_x^2\right)u_{xxx}+u_{xxxxx}=0
\end{align}
which belongs to one of the classes of fifth order integrable equations with constant separant originally found in \cite{Sokolov1985}.  This can again be shown in a similar fashion to the steps performed in (\ref{eq:inttran1})–(\ref{eq:inttran3}). Because of this, the mutually commuting subalgebra containing the generators (\ref{eq:j=-2ell2})–(\ref{eq:j=-2ell5}) is infinite dimensional.
\sm

In \cite{Lindwasser:2024qyh} it was shown that, although these subalgebras look quite different from each other, they are in fact related, forming two groups. In particular, the lowest order Lagrangians (\ref{eq:KdVell1}), (\ref{eq:mKdVell1}), (\ref{eq:j=0ell1}), and (\ref{eq:j=-1ell1}) can all be obtained by taking particular limits of the following order 1 Lagrangian:
\begin{align}
\label{eq:ell12=0}
    \ell_1(j,j_1)=\sqrt{\left(\frac{cj_1}{\sqrt{a(j)}}+b(j)\right)^2+a(j)} + d(j)
\end{align}
where $a(j)$ and $b(j)$ are functionals of $j$ such that $a(j)+b(j)^2$ is a quartic polynomial in $j$, $d(j)$ is a quadratic polynomial functional of $j$, and $c$ is some constant. This is the most general order 1 Lagrangian that commutes with some order 2 Lagrangian $\ell_2$. Meanwhile the lowest order Lagrangians (\ref{eq:BDell2}) and (\ref{eq:j=-2ell2}) can both be obtained by taking particular limits of the following order 2 Lagrangian
\begin{align}
\label{eq:ell23=0}
    \ell_2(j,j_1,j_2)=\big(aj_2+b(j,j_1)\big)^{1/3} +c(j)
\end{align}
where $a$ is a constant, $c(j)$ is a quadratic polynomial functional of $j$, and $b(j,j_1)$ is a cubic polynomial functional of $j_1$,
\begin{align}
\label{eq:bjj1}
    b(j,j_1)=a^{3/2}A(j)j_1^3+aB(j)j_1^2+a^{1/2}C(j)j_1+D(j)
\end{align}
where after making the redefinition
\begin{align}
\label{eq:Bredef}
    B(j)=\frac{C(j)^2- E(j)^2-2D'(j)}{3D(j)}
\end{align}
the functional $E(j)$ is a quadratic polynomial in $j$ and the functional $D(j)$ is a degree six polynomial in $j$, and the prime indicates a derivative with respect to $j$. The functionals $D(j)$ and $E(j)$ also satisfy the coupled equations
\begin{align}
\label{eq:DE1}
&30DD''-25D'^2+4E^4=0 \\
\label{eq:DE2}
&450D^2E''-150DD'E'+25D'^2E-4E^5=0
\end{align}
while $A(j)$ is fully determined in terms of $C(j)$, $D(j)$, and $E(j)$ via a quadratic equation
\begin{align}
\label{eq:A}
    \left(A+\frac{3E^2C-C^3-3CD'+9C'D}{27D^2}\right)^2-\frac{1}{3645\,D^4}\left(2E^3+15DE'-5D'E\right)^2=0
\end{align}
and $C(j)$ is an undetermined functional. This is the most general order 2 Lagrangian that commutes with some order 3 Lagrangian $\ell_3$, but does not commute with an order 1 Lagrangian.
\section{Necessary conditions on Korteweg–de Vries type hierarchies} 
\label{sec:Summary}
In this section, we describe some necessary conditions on the generators $\ell_{n_i}$ of an infinite dimensional mutually commuting subalgebra of (\ref{eq:scalarlie}). We restrict ourselves to consider $1d$ Lagrangians $\ell_{n_i}=\ell_{n_i}(\{j_{n_i}(x^-)\})$, which commute with the energy-momentum tensor $T=j^2$ and hence have no explicit $x^-$ dependence. In particular, we will describe the allowed degeneracies of orders of Lagrangians in the subalgebra, as well as the necessary $j_{n_i}$ dependence of an order $n_i$ Lagrangian $\ell_{n_i}$ in such a subalgebra. Some of the following is a summary of results found in \cite{Lindwasser:2024qyh}.

\sm

We would like to determine some necessary conditions for two $1d$ Lagrangians $\ell_{n_i}$ and $\ell_{n_j}$ with $n_j\geq n_i$ to commute. To commute as in (\ref{eq:KdVcommute}), the equation they must satisfy is 
\begin{align}
\label{eq:ninjcomm}
    \mathcal{E}\left(\partial_-\mathcal{E}(\ell_{n_i})\mathcal{E}(\ell_{n_j})\right) = 0
\end{align}
Equation (\ref{eq:ninjcomm}) depends on $j(x^-)$ and its derivatives up to order $2(n_i+n_j)$. Given that we are looking for commuting Lagrangians $\ell_{n_i}$ and $\ell_{n_j}$ of order $n_i$ and $n_j$, respectively, with $n_j\geq n_i$ regardless of the value of $j(x^-)$ and its derivatives, we may regard this equation as a polynomial equation in the variables $j_{n_j+1},\dots,j_{2(n_i+n_j)}$. To satisfy this equation for any value of $j(x^-)$ and its derivatives, the coefficients of each monomial $j_{n_j+1}^{N_1}\cdots j_{2(n_i+n_j)}^{N_{2n_i+n_j}}$ must vanish. Recall from \cite{Lindwasser:2024qyh} that for Lagrangians with $n_j>n_i>0$, the coefficient of the monomial $j_{2(n_i+n_j)}$ implies
\begin{align}
    (2n_j+1)\partial_-\frac{\partial^2\ell_{n_i}}{\partial j_{n_i}^2}\frac{\partial^2\ell_{n_j}}{\partial j_{n_j}^2}-(2n_i+1)\frac{\partial^2\ell_{n_i}}{\partial j_{n_i}^2}\partial_-\frac{\partial^2\ell_{n_j}}{\partial j_{n_j}^2}=0
\end{align}
Solving for $\ell_{n_j}$, we find that $\ell_{n_j}$ must be quadratic in $j_{n_j}$, taking the form
\begin{align}
\label{eq:ellm}
    \ell_{n_j}(j,\dots,j_{n_j}) = \frac{1}{2}C\left(\frac{\partial^2\ell_{n_i}}{\partial j_{n_i}^2}\right)^{\frac{2n_j+1}{2n_i+1}}j_{n_j}^2 + f_{n_j-1}(j,\dots,j_{n_j-1})
\end{align}
where $C$ is some constant and $f_{n_j-1}$ is some order $n_j-1$ functional. Here we have omitted a possible term linear in $j_{n_j}$, because such a term is always equivalent to a functional of order $n_j-1$ and can be absorbed into $f_{n_j-1}$. Indeed, the total derivative of some order $n_j-1$ functional $g_{n_j-1}(\{j_{n_j-1}(x^-)\})$ is
\begin{align}
    \partial_-g_{n_j-1}=j_{n_j}\frac{\partial g_{n_j-1}}{\partial j_{n_j-1}} + \sum_{k=0}^{n_j-2}j_{k+1}\frac{\partial g_{n_j-1}}{\partial j_k}\sim 0
\end{align}
and so $j_{n_j}\frac{\partial g_{n_j-1}}{\partial j_{n_j-1}}\sim -\sum_{k=0}^{n_j-2}j_{k+1}\frac{\partial g_{n_j-1}}{\partial j_k}$ is equivalent to a functional of order $n_j-1$. In a similar fashion, one can show that any term of the form $j_{n+k}g_{n}$, where $g_n(\{j_n(x^-)\})$ is an order $n$ functional, is equivalent to a functional of order $n+\lfloor k/2\rfloor$. In what follows we will always write a $1d$ Lagrangian $\ell$ in such a way that the maximum order derivative of $j(x^-)$ in each term of $\ell$ cannot be reduced by adding a total derivative.

\sm

This has the important consequence that, in any mutually commuting subalgebra, every generator $\ell_{n_k}$ of order $n_k>0$ must be quadratic in $j_{n_k}$ \textit{except} for the generators with the lowest nonzero order in the mutually commuting subalgebra. 

\sm 

Furthermore when $n_j=n_i=n$, the coefficient of the monomial $j_{2(n_i+n_j)}$ implies that $\ell_{n_j}=\ell_{n}'$ must be proportional to $\ell_{n_i}=\ell_n$ plus some lower order functional
\begin{align}
\label{eq:sameorder}
    \ell_n'(j,\dots,j_n) = C\ell_n(j,\dots,j_n) + f_{n-1}(j,\dots,j_{n-1})
\end{align}
This implies that every mutually commuting subalgebra has a basis of generators such that each generator $\ell_{n_k}$ with $n_k>0$ in this basis has a distinct order. The mutually commuting subalgebras therefore can be organized in the following way. There exists a basis for which there are some order zero generators $\ell_0$, a single generator $\ell_{n_1}$ with lowest nonzero order $n_1>0$, which can be a non-polynomial functional of $j_{n_{1}}$, and then a sequence of generators $\ell_{n_2},\ell_{n_3},\dots,\ell_{n_i},\dots$ of distinct higher orders $n_i>n_1$ and $n_i\neq n_j$, with $\ell_{n_i}$ quadratic in $j_{n_i}$ for $i>1$ as in (\ref{eq:ellm}). Without loss of generality we may take the sequence $n_i$ to be strictly monotonically increasing in $i$. 

\sm

For example, the mutually commuting subalgebras that are so far known to be infinite dimensional are those associated with the Korteweg–de Vries and modified Korteweg–de Vries models, which have generators $\ell_{n_i}$ with $n_1=1$ for all natural numbers $n_i=1,2,3,4,\dots$, i.e. natural numbers satisfying $n_i\equiv 0,1\,\text{(mod }2)$, and the Tzitz\'{e}ica model, which has generators $\ell_{n_i}$ with $n_1=2$ for all natural numbers $n_i$ satisfying $n_i\equiv 0,2\,\text{(mod }3)$. 

\sm

This general structure allows us to characterize each mutually commuting subalgebra, and therefore each Korteweg–de Vries type hierarchy, by its lowest nonzero order generator $\ell_{n_{1}}$. For each value of $n_1$, there is some parameter space of $\ell_{n_1}$, each of which defines a mutually commuting subalgebra. 

\sm

The main purpose of this paper is to determine the necessary functional dependence $\ell_{n_1}$ has on $j_{n_1}$ for each $n_{1}=1,2,\dots$ corresponding to some infinite dimensional mutually commuting subalgebra. To do this, we study the equation
\begin{align}
\label{eq:n1mcomm=0}
     \mathcal{E}\left(\partial_-\mathcal{E}(\ell_{n_1})\mathcal{E}(\ell_m)\right) = 0
\end{align}
We will see that for each $n_1$, and for sufficiently large $m$, the first constraint on $\ell_{n_1}$ comes from the coefficient of the monomial $j_{2m}$ in (\ref{eq:n1mcomm=0}). This constraint on $\ell_{n_1}$ ends up being independent of $m$. Because of the independence of this constraint on sufficiently large $m$, which fixes $\ell_{n_1}$'s $j_{n_1}$ dependence, it must be satisfied if the subalgebra $\ell_{n_1}$ defines is infinite dimensional. We will argue in \autoref{sec:argument} that the coefficient of the monomial $j_{2m}$ implies that $\ell_{n_1}$ must be such that the following quantity is a total derivative
\begin{align}
\label{eq:bigresult}
    \sum_{k=0}^{n_1}\partial_-^{k+1}\mathcal{E}(\ell_{n_1})\frac{\partial}{\partial j_k}\left(\frac{\partial^2\ell_{n_1}}{\partial j_{n_1}^2}\right)^{-\frac{1}{2n_1+1}}
\end{align}

\sm

We have checked that this is true explicitly for $n_1=1,2,3,4$, and it is true more generally for $n_1>4$ depending on some technical assumptions explained in \autoref{sec:argument}. Let us assume that this is true for all $n_1$ and understand its consequences. By adding total derivatives to (\ref{eq:bigresult}), this functional is equivalent to
\begin{align}
     \sum_{k=0}^{n_1}\partial_-^{k+1}\mathcal{E}(\ell_{n_1})\frac{\partial}{\partial j_k}\left(\frac{\partial^2\ell_{n_1}}{\partial j_{n_1}^2}\right)^{-\frac{1}{2n_1+1}}&\sim \partial_-\mathcal{E}(\ell_{n_1})\sum_{k=0}^{n_1}(-1)^k\partial_-^k\frac{\partial}{\partial j_k}\left(\frac{\partial^2\ell_{n_1}}{\partial j_{n_1}^2}\right)^{-\frac{1}{2n_1+1}}\nonumber\\
     &=\partial_-\mathcal{E}(\ell_{n_1})\mathcal{E}\left(\left(\frac{\partial^2\ell_{n_1}}{\partial j_{n_1}^2}\right)^{-\frac{1}{2n_1+1}}\right)\nonumber\\
    &\sim 0
\end{align}
Comparing the expression in the second line with the commutator (\ref{eq:scalarlie}), we see that this constraint means that $\ell_{n_1}$ commutes with $(\ell_{n_1}'')^{-1/(2n_1+1)}$, where the primes indicate derivatives with respect to $j_{n_1}$. This is related to the fact that any integrable equation of the form (\ref{eq:quasilinearV}) with a potential $V(\{u,u_x,\dots,u_{2n}\})$ that is order $2n$ has a conserved current $J_i$ with time component $J_t=\rho_{-1}$ \cite{Sokolov1984class},
\begin{align}
    \rho_{-1}=\left(\frac{\partial V}{\partial u_{2n}}\right)^{-1/(2n+1)}
\end{align}
Instead, we have discovered that in any infinite dimensional mutually commuting subalgebra with lowest nonzero order generator $\ell_{n_1}$, $(\ell_{n_1}'')^{-1/(2n_1+1)}$ commutes with $\ell_{n_1}$, and we will see shortly that it is in fact an element of the subalgebra. Because of this, Korteweg–de Vries type hierarchies derived from (\ref{eq:scalarlambdaaction}) have a conserved current with time component $(\ell_{n_1}'')^{-1/(2n_1+1)}$, which after converting to the conventions (\ref{eq:KdVconv}) is the same as $\rho_{-1}$. We have shown, however, that this is a property of \textit{any} integrable deformation $\delta_{\ell_{n_i}}S_{\text{deform}}=0$ of the $2d$ single scalar conformal field theory.

\sm

Let us work out what this means with the additional input that $\ell_{n_1}$ is the lowest nonzero order generator in the subalgebra. Assuming first that $\ell_{n_1}'''\neq 0$, $(\ell_{n_1}'')^{-1/(2n_1+1)}$ is a $1d$ Lagrangian of order $n_1$. Recall that, when two Lagrangians of the same order commute, they are related via (\ref{eq:sameorder}), and so 
\begin{align}
\label{eq:secondorderelln1}
    (\ell_{n_1}'')^{-1/(2n_1+1)}=C\ell_{n_1}+f_{n_1-1}
\end{align}
where $C$ is some constant, and $f_{n_1-1}$ is some order $n_1-1$ functional to be determined. Because $\ell_{n_1}$ commutes with $(\ell_{n_1}'')^{-1/(2n_1+1)}$, (\ref{eq:secondorderelln1}) implies that $\ell_{n_1}$ also commutes with $f_{n_1-1}$. But we assumed $\ell_{n_1}$ has the lowest nonzero order in the mutually commuting subalgebra, and $f_{n_1-1}$ has a lower order than $\ell_{n_1}$, so $f_{n_1-1}=f(j)$ must be an order 0 functional. Without loss of generality, we may remove $f(j)$ from (\ref{eq:secondorderelln1}) by redefining $\ell_{n_1}\to\ell_{n_1}-f(j)/C$,
\begin{align}
\label{eq:secondordersimpl}
    (\ell_{n_1}'')^{-1/(2n_1+1)}=C\ell_{n_1}
\end{align}
The functional $f(j)$ simply expresses the freedom we have in adding to $\ell_{n_1}$ any order 0 Lagrangian in the subalgebra. Because $(\ell_{n_1}'')^{-1/(2n_1+1)}$ is proportional $\ell_{n_1}$, it commutes not just with $\ell_{n_1}$, but with every generator $\ell_{n_i}$ in the subalgebra. Solving (\ref{eq:secondordersimpl}), we get an implicit solution for $\ell_{n_1}$, which after redefining variables takes the form
\begin{align}
\label{eq:transcend}
    \left(\ell_{n_1}{}_2F_1\left(\frac{1}{2},-\frac{1}{2n_1};1-\frac{1}{2n_1};a_{n_1-1}(\ell_{n_1})^{-2n_1}\right)\right)^2=\left(\frac{cj_{n_1}}{\sqrt{a_{n_1-1}}}+b_{n_1-1}\right)^2
\end{align}
where now $a_{n_1-1}$ and $b_{n_1-1}$ are order $n_1-1$ functionals that come from integration constants when solving (\ref{eq:secondordersimpl}), and $c\neq 0$ is some constant independent of all $j_k$, related to $C$.

\sm

When $n_1=1$, (\ref{eq:transcend}) can be solved analytically for $\ell_1$,
\begin{align}
\label{eq:ell1nec}
    \ell_1(j,j_1) = \sqrt{\left(\frac{cj_1}{\sqrt{a(j)}}+b(j)\right)^2+a(j)}
\end{align}
Recall from \cite{Lindwasser:2024qyh} that the space of order 1 Lagrangians $\ell_1$ that commute with an order 2 Lagrangian $\ell_2$ takes precisely this form up to some order 0 functional, with the additional constraint that $a(j)+b(j)^2$ is a quartic polynomial in $j$.

\sm

When $n_1>1$, (\ref{eq:transcend}) cannot be solved for $\ell_{n_1}$ analytically for all choices of $a_{n_1-1}$, $b_{n_1-1}$ and $c$. However, an analytic solution for $\ell_{n_1}$ can be obtained for special choices of parameters after redefining variables,
\begin{align}
   & |c|\to\frac{1}{n_1+1}|a|,&& b_{n_1-1}\to\frac{b_{n_1-1}}{(n_1+1)\sqrt{-a_{n_1-1}}}+\frac{\Gamma(1-\frac{1}{2n_1})\Gamma(\frac{1}{2}+\frac{1}{2n_1})}{\sqrt{\pi}}(-a_{n_1-1})^{\frac{1}{2n_1}}
\end{align}
with $a$ some new constant, and the phase of $a$ is chosen so that in the asymptotic limit $|a_{n_1-1}|\to\infty$, $\ell_{n_1}$ equals
\begin{align}
\label{eq:ellnnec}
    \ell_{n_1}(j,\dots,j_{n_1}) = \left(aj_{n_1}+b_{n_1-1}(j,\dots,j_{n_1-1})\right)^{\frac{1}{n_1+1}}
\end{align}
Recall from \cite{Lindwasser:2024qyh} that the space of order 2 Lagrangians $\ell_{2}$ that commute with an order 3 Lagrangian $\ell_3$ but not an order 1 Lagrangian is up to some order 0 functional
\begin{align}
    \ell_2(j,j_1,j_2) = \big(aj_2+b_1(j,j_1)\big)^{1/3}
\end{align}
where $b_1(j,j_1)$ is a cubic polynomial in $j_1$. This is precisely of the form (\ref{eq:ellnnec}), with some additional constraints on the lower order functional $b_{n_1-1}$, reviewed in \autoref{ssec:mutcommutsub}.

\sm

Finally if we assume $\ell_{n_1}'''=0$, or in other words if $\ell_{n_1}$ is quadratic in $j_{n_1}$, then the functional $(\ell_{n_1}'')^{-1/(2n_1+1)}$ is at most order $n_1-1$. But because it commutes with $\ell_{n_1}$, which is assumed to be the lowest nonzero order Lagrangian in the mutually commuting subalgebra, $(\ell_{n_1}'')^{-1/(2n_1+1)}=f(j)$ must be an order 0 functional. Barring some exceptional coincidences discussed in \cite{Lindwasser:2024qyh}, the order 0 generator $f(j)$ in any mutually commuting subalgebra consisting of generators without explicit $x^-$ dependence is a quadratic functional of $j$, and so up to some normalization
\begin{align}
    \ell_{n_1}=\frac{1}{2}\frac{j_{n_1}^2}{(aj^2+bj+c)^{2n_1+1}}+f_{n_1-1}
\end{align}
with $f_{n_1-1}$ some undetermined order $n_1-1$ functional, and $a,b,c$ are constants. Using (\ref{eq:ellm}), this is precisely the functional dependence of all other generators $\ell_{n_i}$ with $n_i>n_1$ in the subalgebra after replacing $n_1\to n_i$,
\begin{align}
\label{eq:quadcomm}
    \ell_{n_i}=\frac{1}{2}\frac{j_{n_i}^2}{(aj^2+bj+c)^{2n_i+1}}+f_{n_i-1}
\end{align}

\sm

By demanding that $[j]=0$, and searching for mutually commuting subalgebras with definite scaling dimension Lagrangians, the Lagrangians must take the form (\ref{eq:quadcomm}), as is the case for the subalgebra with $n_1=1$ with the first three generators (\ref{eq:j=0ell1})–(\ref{eq:j=0ell3}). Because $\ell_{n_1}$ is quadratic in $j_{n_1}$, it still has a chance of commuting with a Lagrangian with nonzero order lower than $n_1$. For instance, by demanding that an order 2 Lagrangian $\ell_2$ and order 3 Lagrangian of the form (\ref{eq:quadcomm}) commute, one gets automatically that both also commute with an order 1 Lagrangian $\ell_1$. This means that there are no mutually commuting subalgebras with generators of the form (\ref{eq:quadcomm}) with a sequence of orders $n_i$ such that $n_1=2$ and $n_2=3$.
\section{$\mathcal{E}(\partial_-\mathcal{E}(\ell_{n_1})\mathcal{E}(\ell_m))=0$}
\label{sec:argument}
In this section, we will study the equation (\ref{eq:n1mcomm=0}) for any $m>n_1>0$ until we find a constraint on $\ell_{n_1}$ that must be satisfied for any $m\geq C$, with some constant $C>0$. Such a constraint must be satisfied if the mutually commuting subalgebra $\ell_{n_1}$ defines is infinite dimensional. We will argue in this section that as long as $m\geq 4n_1-2$, $\ell_{n_1}$ and $\ell_m$ commuting requires that the quantity (\ref{eq:bigresult}) is a total derivative.

\sm

To show this, we will derive a recursion relation from (\ref{eq:n1mcomm=0}) which solves for certain components $B_{m-i}$ of $\ell_m$ in terms of $\ell_{n_1}$, to be defined shortly. Recall that we assume that $\ell_m$ is a local functional of $j(x^-)$ and its derivatives of maximum order $m$. Crucially for our argument, the resulting recursion relation cannot preserve this assumption unless $\ell_{n_1}$ satisfies certain conditions.

\sm

To begin, recall that $\mathcal{E}(\partial_-\mathcal{E}(\ell_{n_1})\mathcal{E}(\ell_m))=0$ is a polynomial equation that involves $j(x^-)$ and its derivatives up to order $2(n_1+m)$. So if $\mathcal{E}(\partial_-\mathcal{E}(\ell_{n_1})\mathcal{E}(\ell_m))=0$, then any derivative of $\mathcal{E}(\partial_-\mathcal{E}(\ell_{n_1})\mathcal{E}(\ell_m))$, say with respect to $j_{2(n_1+m)-p}$ with $0\leq p\leq 2(n_1+m)$, is zero
\begin{align}
\label{eq:pn1mcomm=0}
    \frac{\partial}{\partial j_{2(n_1+m)-p}}\mathcal{E}(\partial_-\mathcal{E}(\ell_{n_1})\mathcal{E}(\ell_m))=0
\end{align}
Because $\mathcal{E}(\partial_-\mathcal{E}(\ell_{n_1})\mathcal{E}(\ell_m))$ can be written as a commutator of infinitesimal symmetry transformations $4[\delta_{\ell_{n_1}},\delta_{\ell_m}]\phi$, we may write $\mathcal{E}(\partial_-\mathcal{E}(\ell_{n_1})\mathcal{E}(\ell_m))$ using (\ref{eq:scalarliedirect}) as
\begin{align}
    \mathcal{E}(\partial_-\mathcal{E}(\ell_{n_1})\mathcal{E}(\ell_m))=4[\delta_{\ell_{n_1}},\delta_{\ell_m}]\phi=\sum_{k=0}^{2m}\partial_-^{k+1}\mathcal{E}(\ell_{n_1})\frac{\partial}{\partial j_k}\mathcal{E}(\ell_m)-\sum_{k=0}^{2n_1}\partial_-^{k+1}\mathcal{E}(\ell_m)\frac{\partial}{\partial j_k}\mathcal{E}(\ell_{n_1})
\end{align}
We first focus on (\ref{eq:pn1mcomm=0}) when $0\leq p < 2n_1$. Using the above and the commutation relation $[\frac{\partial}{\partial j_k},\partial_-]=\frac{\partial}{\partial j_{k-1}}$, (\ref{eq:pn1mcomm=0}) becomes
\begin{align}
\label{eq:recursion1}
    &\frac{\partial}{\partial j_{2(n_1+m)-p}}\mathcal{E}(\partial_-\mathcal{E}(\ell_{n_1})\mathcal{E}(\ell_m)) = \nonumber\\
    &\hspace{-.8cm}+\sum_{k=0}^p\sum_{r=0}^k\Bigg(\hspace{-0.09cm}{2m-p+1+k \choose k+1-r}\partial_-^{k+1-r}\frac{\partial \mathcal{E}(\ell_{n_1})}{\partial j_{2n_1-r}}\frac{\partial\mathcal{E}(\ell_m)}{\partial j_{2m-p+k}}-{2n_1+1-r\choose k+1-r}\partial_-^{k+1-r}\frac{\partial \mathcal{E}(\ell_m)}{\partial j_{2m-p+k}}\frac{\partial\mathcal{E}(\ell_{n_1})}{\partial j_{2n_1-r}}\Bigg) \nonumber\\
    &=0
\end{align}
For $0\leq p < 2n_1$, this is a recursion relation for $\frac{\partial \mathcal{E}(\ell_m)}{\partial j_{2m-p}}$ of order $p$, which can be used to solve for $\frac{\partial \mathcal{E}(\ell_m)}{\partial j_{2m-p}}$ in terms of $\frac{\partial \mathcal{E}(\ell_{n_1})}{\partial j_{2n_1-r}}$ with $0\leq r\leq p$.

\sm

We simplify the recursion relation further by expanding out the derivatives $\frac{\partial \mathcal{E}(\ell_{n_1})}{\partial j_{2n_1-r}}$ and $\frac{\partial\mathcal{E}(\ell_m)}{\partial j_{2m-r}}$, and noting that they can be written in the following form
\begin{align}
\label{eq:dEexpans}
    &\frac{\partial\mathcal{E}(\ell_{n_1})}{\partial j_{2n_1-r}}=\sum_{i=0}^{\lfloor r/2\rfloor}{n_1-i\choose r-2i}\partial^{r-2i}_-A_{n_1-i}, && \frac{\partial\mathcal{E}(\ell_m)}{\partial j_{2m-r}}=\sum_{i=0}^{\lfloor r/2\rfloor}{m-i\choose r-2i}\partial^{r-2i}_-B_{m-i}
\end{align}
where $A_{n_1-i}$ and $B_{m-i}$ are functionals of $\ell_{n_1}$ and $\ell_m$, respectively. The binomial coefficients ${n\choose k}$ in the above expressions are understood to vanish when $k>n$. The explicit functional expression for $B_{m-i}$ is
\begin{align}
\label{eq:Bm-idef}
    (-1)^{m-i}B_{m-i}=&\frac{\partial^2\ell_m}{\partial j_{m-i}^2}+2\sum_{k=1}^i(-1)^k\frac{\partial^2\ell_m}{\partial j_{m-i+k}\partial j_{m-i-k}}+\sum_{k=1}^{i}(-1)^k\partial_-^k\frac{\partial^2\ell_m}{\partial j_{m-i+k}\partial j_{m-i}}\nonumber\\
    &+\sum_{k=1}^i\sum_{r=1}^{i-k}(-1)^{k+r}\left[{k+r-1\choose r}+2{k+r-2\choose r-1}\right]\partial_-^k\frac{\partial^2\ell_m}{\partial j_{m-i+k+r}\partial j_{m-i-r}}
\end{align}
and likewise for $A_{n_1-i}$ after the replacements $m\to n_1$ and $\ell_m\to\ell_{n_1}$, with the understanding that in the above any derivative with respect to $j_k$ with $k<0$ is zero. The components $B_{m-i}$ of $\ell_{m}$ are such that the only non-mixed second order derivative present is, apart from $x^-$, with respect to $j_{m-i}$, and that $B_{m-i}=0$ if $\ell_m=\partial_-f_{m-1}$ is a total derivative of an order $m-1$ functional $f_{m-1}$, and likewise for $A_{n_1-i}$. We list here the first couple $A_{n_1-i}$ and $B_{m-i}$ in terms of $\ell_{n_1}$ and $\ell_m$,
\begin{align}
\label{eq:AnBmdef}
    &A_{n_1}=(-1)^{n_1}\frac{\partial^2\ell_{n_1}}{\partial j_{n_1}^2}, && B_m=(-1)^m\frac{\partial^2\ell_{m}}{\partial j_{m}^2}
\end{align}
In terms of $A_{n_1-i}$ and $B_{m-i}$, the recursion relation (\ref{eq:recursion1}) when $p=2q<2n_1$ becomes
\begin{align}
\label{eq:recursion2}
    \sum_{i=0}^q\sum_{j=0}^i\sum_{k=0}^{2(i-j)}\Big(a^q_{ijk}\partial_-^{k+1}B_{m-q+i}\partial_-^{2(i-j)-k}A_{n_1-j}-b^q_{ijk}\partial_-^{k+1}A_{n_1-j}\partial_-^{2(i-j)-k}B_{m-q+i}\Big)=0
\end{align}
where $a^q_{ijk}$ and $b^q_{ijk}$ are coefficients equal to
\begin{align}
    a^q_{ijk} & = {n_1-j\choose 2(i-j)-k}\Bigg[{2n_1+m-q-i+k+1\choose k+1 }-{m-q+i\choose k+1}\Bigg]\\
    b^q_{ijk} &= {m-q+i\choose 2(i-j)-k}\Bigg[{2(m-q)+n_1+j+k+1\choose k+1}-{n_1-j\choose k+1}\Bigg]
\end{align}
Now for $0\leq q < n_1$, this is a recursion relation for $B_{m-q}$ of order $q$, which can be used to solve for $B_{m-q}$ in terms of $A_{n_1-i}$ with $0\leq i\leq q$. Isolating $B_{m-q}$ in (\ref{eq:recursion2}), we find
\begin{align}
\label{eq:recursion3}
    &-(2n_1+1)A_{n_1}^{\frac{2(m+n_1-q+1)}{2n_1+1}}\partial_-\left(\frac{B_{m-q}}{A_{n_1}^{\frac{2(m-q)+1}{2n_1+1}}}\right) \nonumber\\
    &=\sum_{i=1}^q\sum_{j=0}^i\sum_{k=0}^{2(i-j)}\Big(a^q_{ijk}\partial_-^{k+1}B_{m-q+i}\partial_-^{2(i-j)-k}A_{n_1-j}-b^q_{ijk}\partial_-^{k+1}A_{n_1-j}\partial_-^{2(i-j)-k}B_{m-q+i}\Big)
\end{align}

\sm

Let us try to understand the structure of this recursion relation in detail. In order for $B_{m-q}$, related to $\ell_m$ via (\ref{eq:Bm-idef}), to be a local functional of $j(x^-)$ and its derivatives, the right-hand side of (\ref{eq:recursion3}) multiplied by $A_{n_1}^{-\frac{2(m+n_1-q+1)}{2n_1+1}}$ must be a total derivative. Once this is the case, this equation can be freely integrated to solve for $B_{m-q}$, which is unique up to some integration constant. These integration constants, coming from solving for all $B_{m-i}$ up to $i=q$, amount to $\ell_m$ being written as a linear combination of functionals of lower order 
\begin{align}
    \ell_m = c_0\tilde{\ell}_m + c_1\tilde{\ell}_{m-1} + \cdots + c_q\tilde{\ell}_{m-q}
\end{align}
for which each $\tilde{\ell}_{m-i}$ commutes with $\ell_{n_1}$ independent of the others. Without loss of generality then, we may set $c_0=(-1)^{\frac{m-n_1}{2n_1+1}}$ and $c_i=0$ when $i\geq 1$ because a $c_i\neq 0$ amounts to solving for a functional $\ell_{m-i}$ of lower order that commutes with $\ell_{n_1}$. With this convention, $B_m$ can, for instance, be solved from (\ref{eq:recursion3}) with $q=0$,
\begin{align}
\label{eq:Bm}
    B_{m}=(-1)^{\frac{m-n_1}{2n_1+1}}A_{n_1}^{\frac{2m+1}{2n_1+1}}=(-1)^m\left(\frac{\partial^2\ell_{n_1}}{\partial j_{n_1}^2}\right)^{\frac{2m+1}{2n_1+1}}
\end{align}
which upon integrating twice with respect to $j_m$ reproduces (\ref{eq:ellm}) with $C=1$.

\sm

We find by directly solving for $B_{m-q}$ up to $q=4$ that the right-hand side of (\ref{eq:recursion3}) multiplied by $A_{n_1}^{-\frac{2(m+n_1-q+1)}{2n_1+1}}$ is, in fact, a total derivative, regardless of the values of $A_{n_1-i}$ with $0\leq i\leq q$ and therefore $\ell_{n_1}$. This is a nontrivial and important property of this recursion relation, depending sensitively on the coefficients $a^q_{ijk}$ and $b^q_{ijk}$, which at this stage we cannot prove for general $0\leq q \leq n_1$ for $n_1>4$, but we simply assume is true hereafter. We discuss more details of this property later in \autoref{sec:discussion}. If this were not true, then this would imply a constraint on the $A_{n_1-i}$'s and therefore $\ell_{n_1}$ in order for the $B_{m-q}$'s to be local functionals of $j(x^-)$ and its derivatives.

\sm

Basic derivative counting of this recursion relation shows that $B_{m-q} = (-1)^{m-q}C_{n_1+2q}$, where $C_{n_1+2q}$ is a complicated functional of $j(x^-)$ and its derivatives up to order $n_1+2q$ depending on $\ell_{n_1}$ for all $0\leq q < n_1$. We show $C_{n_1}$ and $C_{n_1+2}$ explicitly in (\ref{eq:Cn}) and (\ref{eq:Cn+2}), respectively, and $C_{n_1+4},C_{n_1+6},C_{n_1+8}$ in the \textit{Mathematica} ancillary file \texttt{Bm-q.m}. Once $C_{n_1+2q}$ is obtained, the equation $B_{m-q} = (-1)^{m-q}C_{n_1+2q}$ can be used to solve partially for $\ell_m$ using (\ref{eq:Bm-idef}). As long as $m - q > n_1 + 2q$, or $0\leq q \leq \lfloor\frac{m-n_1-1}{3}\rfloor$, one can show recursively that $B_{m-q}$ in (\ref{eq:Bm-idef}) simplifies to
\begin{align}
\label{eq:Bm-qsimple}
    B_{m-q} = (-1)^{m-q}\frac{\partial^2\ell_m}{\partial j_{m-q}^2}=(-1)^{m-q}C_{n_1+2q}
\end{align}
Given that the recursion relation (\ref{eq:recursion3}) is only valid for $0\leq q < n_1$, different things happen depending on whether $\lfloor\frac{m-n_1-1}{3}\rfloor + 1\geq n_1$ or $\lfloor\frac{m-n_1-1}{3}\rfloor + 1< n_1$. We will ignore the latter case, because it is only true for all $n_1$ when $m\leq 4n_1 -3$, in which case there does not necessarily exist a constant $C$ such that whatever constraints on $\ell_{n_1}$ derived in this case apply for any $m\geq C$.

\sm

Consider then the case when $\lfloor\frac{m-n_1-1}{3}\rfloor + 1\geq n_1$, which is satisfied when $m\geq 4n_1-2$. In this case, $B_{m-q}$ is (\ref{eq:Bm-qsimple}) for all $0\leq q <n_1$, and this fixes the $j_{m-q}$ dependence of $\ell_m$ by integrating (\ref{eq:Bm-qsimple}) twice with respect to $j_{m-q}$,
\begin{align}
\label{eq:ellmwhenm-n-1/3<n}
    \ell_m = \frac{1}{2}\sum_{q=0}^{n_1-1}C_{n_1+2q}j_{m-q}^2 + f_{m-n_1}
\end{align}
where $f_{m-n_1}$ is some undetermined functional of order $m-n_1$. Here we have omitted in $\ell_m$ any term linear in $j_{m-q}$, because such a term is equivalent to a functional of order $m-q-1$.

\sm

Therefore, no constraints on $\ell_{n_1}$ are found at this stage. Instead when $m\geq 4n_1-2$, the equations (\ref{eq:pn1mcomm=0}) for $0\leq p< 2n_1$ only fix the functional dependence of $\ell_{m}$ on $j_{m-q}$ for $0\leq q < n_1$.

\sm

When $p = 2n_1$, which corresponds to the coefficient of the monomial $j_{2m}$ of the equation $\mathcal{E}(\partial_-\mathcal{E}(\ell_{n_1})\mathcal{E}(\ell_m))=0$, we see our first constraint on $\ell_{n_1}$. In this case, (\ref{eq:pn1mcomm=0}) gives an equation for $B_{m-n_1}$
\begin{align}
\label{eq:recursion4}
    &-(2n_1+1)A_{n_1}^{\frac{2(m+1)}{2n_1+1}}\partial_-\left(\frac{B_{m-n_1}}{A_{n_1}^{\frac{2(m-n_1)+1}{2n_1+1}}}\right) = \sum_{k=0}^{2m}\partial_-^{k+1}\mathcal{E}(\ell_{n_1})\frac{\partial}{\partial j_k}\frac{\partial\mathcal{E}(\ell_m)}{\partial j_{2m}} \nonumber\\
    &+\sum_{i=1}^{n_1}\sum_{j=0}^i\sum_{k=0}^{2(i-j)}\Big(a^{n_1}_{ijk}\partial_-^{k+1}B_{m-n_1+i}\partial_-^{2(i-j)-k}A_{n_1-j}-b^{n_1}_{ijk}\partial_-^{k+1}A_{n_1-j}\partial_-^{2(i-j)-k}B_{m-n_1+i}\Big)
\end{align}
Note that this differs from (\ref{eq:recursion3}) only by the addition of the term on the right-hand side of the first line. If the recursion relation (\ref{eq:recursion3}) satisfies the property that the right-hand side multiplied by $A_{n_1}^{-\frac{2(m+n_1-q+1)}{2n_1+1}}$ is a total derivative, regardless of the values of $A_{n_1-i}$ with $0\leq i\leq q$ up to and including $q=n_1$, then consistency of (\ref{eq:recursion4}) requires that the additional term multiplied by $A_{n_1}^{-\frac{2(m+1)}{2n_1+1}}$ is also a total derivative, i.e. after using (\ref{eq:dEexpans}), (\ref{eq:Bm}) and (\ref{eq:AnBmdef}), the quantity
\begin{align}
\label{eq:bigresultsec5}
    \sum_{k=0}^{n_1}\partial_-^{k+1}\mathcal{E}(\ell_{n_1})\frac{\partial}{\partial j_k}\left(\frac{\partial^2\ell_{n_1}}{\partial j_{n_1}^2}\right)^{-\frac{1}{2n_1+1}}
\end{align}
is a total derivative. The consequences of this were worked out in \autoref{sec:Summary}.
\section{Discussion}
\label{sec:discussion}
In this paper, we found necessary conditions (\ref{eq:transcend}) on all possible infinite dimensional mutually commuting subalgebras of the $2d$ single scalar conformal field theory. In particular, we found the necessary $j_{n_1}$ dependence of the lowest order Lagrangian $\ell_{n_1}$ of any infinite dimensional mutually commuting subalgebra of (\ref{eq:scalarlie}) when $n_1=1,2,3,4$ and more generally for $n_1>4$ if the recursion relation (\ref{eq:recursion3}) satisfies the property that the right-hand side multiplied by $A_{n_1}^{-\frac{2(m+n_1-q+1)}{2n_1+1}}$ is a total derivative regardless of the values of all $A_{n_1-i}$, for all $0\leq q\leq n_1$. 

\sm

We explicitly verified that this property is true for $q=0,1,2,3,4$ in (\ref{eq:recursion3}) by recursively solving for $B_{m-q}$ and checking. We list the first couple $B_{m-q}$ here
\begin{align}
\label{eq:Cn}
    &(-1)^{\frac{n_1-m}{2n_1+1}}B_m=A_{n_1}^{\frac{2m+1}{2n_1+1}}\\
\label{eq:Cn+2}
    &(-1)^{\frac{n_1-m}{2n_1+1}}B_{m-1}=\frac{2m+1}{2n_1+1}A_{n_1}^{\frac{2(m-n_1)}{2n_1+1}}A_{n_1-1}+\frac{1}{6}(m-n_1)(m+n_1+1)\frac{2m+1}{2n_1+1}A_{n_1}^{\frac{2(m-n_1)}{2n_1+1}}A_{n_1}''\nonumber\\
    &\hspace{2.65cm}-\frac{1}{6}(m-n_1)(m+n_1+1)\frac{(2m+1)(2n_1-1)}{(2n_1+1)^2}A_{n_1}^{\frac{2m-4n_1-1}{2n_1+1}}(A_{n_1}')^2
\end{align}
where the primes here indicate an $x^-$ derivative. 
For $q=2,3,4$, $B_{m-q}$ are more complicated, and so we provide them in the \textit{Mathematica} ancillary file \texttt{Bm-q.m}. When $q>4$, the right-hand side is sufficiently complicated so that it becomes too time consuming to check that this property is satisfied recursively.

\sm

To prove that this property is true for all $0\leq q \leq n_1$ when $n_1>4$, other methods are necessary. Mathematical induction seems to be the most appropriate method, although for reasons we explain below there are some technical challenges to doing so. In particular, we may phrase this property in the following way. Because $B_{m-q}$ is a functional of $j(x^-)$ and its derivatives only through its dependence on $A_{n_1-i}$ with $0\leq i\leq q$, we define an Euler operator $\mathcal{E}^{n_1-i}$ with respect to these variables
\begin{align}
    \mathcal{E}^{n_1-i}(f)\equiv\sum_{k=0}^\infty(-1)^k\partial_-^k\frac{\partial}{\partial(\partial_-^kA_{n_1-i})}f
\end{align}
That the right-hand side of (\ref{eq:recursion3}) multiplied by $A_{n_1}^{-\frac{2(m+n_1-q+1)}{2n_1+1}}$, which we will call $D_q$ as a shorthand, is a total derivative then means that
\begin{align}
\label{eq:induction}
    &\mathcal{E}^{n_1-i}(D_q)=0,&& \forall\,0\leq i\leq q
\end{align}
Given that we already have a base case, proving that this is true for all $0\leq q\leq n_1$ by mathematical induction would amount to proving that if (\ref{eq:induction}) is true for all $0\leq q\leq n-1$ given that $B_{m-j}$ satisfies the recursion relation (\ref{eq:recursion3}) for all $0\leq j\leq n-1$, then it is also true at $q=n$. For example, checking this at $i=q=n$, we find
\begin{align}
    \mathcal{E}^{n_1-q}(D_q)=&a^q_{qq0}\partial_-B_mA_{n_1}^{-\frac{2(m+n_1-q+1)}{2n_1+1}}+b^q_{qq0}\partial_-\Big(B_mA_{n_1}^{-\frac{2(m+n_1-q+1)}{2n_1+1}}\Big)\nonumber\\
    =&2(m+n_1-q+1)A_{n_1}^{-1-\frac{2(m+n_1-q+1)}{2n_1+1}}\Big(\partial_-B_mA_{n_1}-\frac{2m+1}{2n_1+1}B_m\partial_-A_{n_1}\Big)
\end{align}
which is zero by virtue of (\ref{eq:recursion3}) at $q=0$. Because of the form of the recursion relation (\ref{eq:recursion3}), checking that $\mathcal{E}^{n_1-i}(D_q)=0$ at $i<q=n$ will require more and more input from (\ref{eq:recursion3}) as $i$ gets smaller. In particular, checking that $\mathcal{E}^{n_1-i}(D_q)=0$ requires knowledge of $B_{m-j}$ for all $0\leq j \leq q-i$ when $1\leq i \leq q$, and for all $0\leq j\leq q-1$ when $i=0$. This makes it difficult to prove by mathematical induction. Without a more promising direction for proving this property at this time, we will simply leave it here as a conjecture.

\sm

Beyond these technical considerations, there are several open questions worth exploring in the future. For instance, while we have determined the necessary $j_{n_1}$ dependence of $\ell_{n_1}$, it will be interesting to determine what, if any, necessary constraints there are on the undetermined functionals $a_{n_1-1}(j,\dots,j_{n_1-1})$ and $b_{n_1-1}(j,\dots,j_{n_1-1})$ in (\ref{eq:transcend}). As we saw in \cite{Lindwasser:2024qyh}, there are further constraints on $a_{n_1-1}$ and $b_{n_1-1}$ if one requires that $\ell_{n_1}$ commutes with $\ell_m$ of a given order $m$. However, there may be other subalgebras with a sequence of Lagrangians of orders $n_1,n_2,\dots,n_i,\dots$ with the same $n_1$ but all $n_i\neq m$. What we would like to find then are constraints on $a_{n_1-1}$ and $b_{n_1-1}$ that are independent of $m$, as we did for the $j_{n_1}$ dependence of $\ell_{n_1}$. To do so, it is necessary to impose (\ref{eq:pn1mcomm=0}) when $p>2n_1$.

\sm

Given an infinite dimensional mutually commuting subalgebra with lowest nonzero order Lagrangian $\ell_{n_1}$, it would be interesting to determine what sequences $n_1,n_2,\dots,n_i,\dots$ of orders of Lagrangians are allowed. Recall that when $n_1=1$, all subalgebras found obey $n_i\equiv 0,1\,(\text{mod 2})$, whereas when $n_1=2$, all subalgebras found obey $n_i\equiv 0,2\,(\text{mod 3})$. With the limited data we have, it appears there is a unique sequence of orders for each $n_1$. It is tempting to propose that this is true for all $n_1$, and that the sequence obeys $n_i\equiv 0,n_1\,(\text{mod }n_2)$. Whether or not this is true would be interesting to pursue in the future. To begin with one should at least find subalgebras with $n_1>2$, perhaps by using the ansatz $\ell_{n_1}=(1+j_{n_1})^{\frac{1}{n_1+1}}$ to search for commuting Lagrangians with definite scaling dimension assuming $[j]=-n_1$ and $[x^-]=-1$. Interestingly, it was proven through other means that there are no infinite dimensional mutually commuting subalgebras of definite scaling dimension polynomial Lagrangians with $n_1>2$ when  $[j]>0$ \cite{SANDERS1998410}. We would like to understand if this is true more generally in the future.

\sm

The methods presented in this paper are best suited for \textit{narrowing in} on the space of integrable deformations. They cannot at present prove that any given mutually commuting subalgebra is infinite dimensional. It may, in fact, be the case that a subalgebra defined by $\ell_{n_1}$ for a given $a_{n_1-1}$, $b_{n_1-1}$, and $c$ is either not infinite dimensional, or does not exist. Paired with these methods, we would like to develop methods of proving integrability of the models found in this paper.

\sm

The same analysis can be done for any theory with left- (right-) moving currents, such as a collection of scalars, a collection of left- (right-) moving fermions, Wess–Zumino–Witten models, or principal chiral models. All of these theories have an infinite dimensional Lie algebra, whose infinite dimensional mutually commuting subalgebras define integrable hierarchies, which will be interesting to find. Finally, there is a wealth of literature on integrable deformations \cite{Klimcik:2008eq,Klimcik:2014bta,Sfetsos:2013wia,Zarembo:2017muf,Orlando:2019his,Klimcik:2021bjy,Hoare:2021dix,Ferko:2024ali,Bielli:2024fnp} of these models that are not of the type we considered here, and it will be interesting to determine if the methods in this paper are related, or can perhaps be accommodated to reproduce, these integrable deformations.

\subsection*{Acknowledgments}
The author would like to thank Christian Ferko and Alexander V. Mikhailov for useful discussions, and the Center for Cosmology and Particle Physics at New York University for their hospitality during the preparation of this work. The author is supported by the Taiwan NSTC Grant No. 113-2811-M-002 -167 -MY3 and the Yushan Young Fellowship.

\bibliography{KdVHierarchies}

\providecommand{\href}[2]{#2}\begingroup\raggedright\begin{thebibliography}{10}

\bibitem{Boussinesq}
J.V.~Boussinesq, \emph{Theorie de i’intumescence liquid, appleteonde solitaire au de translation, se propageantdansun canal rectangulaire}, {\emph{Les Comptes Rendus de l'Académie des Sciences} {\bfseries 72} (1871) 755}.

\bibitem{russell1837experimental}
J.S.~Russell, \emph{Experimental researches into the laws of certain hydrodynamical phenomena that accompany the motion of floating bodies and have not previously been reduced into conformity with the known laws of the resistance of fluids}, Royal Society of Edinburgh (1837).

\bibitem{Korteweg01051895}
D.J.~Korteweg and G.~de~Vries, \emph{On the change of form of long waves advancing in a rectangular canal, and on a new type of long stationary waves}, \href{https://doi.org/10.1080/14786449508620739}{\emph{The London, Edinburgh, and Dublin Philosophical Magazine and Journal of Science} {\bfseries 39} (1895) 422}.

\bibitem{PhysRevLett.15.240}
N.J.~Zabusky and M.D.~Kruskal, \emph{Interaction of ``solitons" in a collisionless plasma and the recurrence of initial states}, \href{https://doi.org/10.1103/PhysRevLett.15.240}{\emph{Phys. Rev. Lett.} {\bfseries 15} (1965) 240}.

\bibitem{Gardner:1967wc}
C.S.~Gardner, J.M.~Greene, M.D.~Kruskal and R.M.~Miura, \emph{{Method for solving the Korteweg-de Vries equation}}, \href{https://doi.org/10.1103/PhysRevLett.19.1095}{\emph{Phys. Rev. Lett.} {\bfseries 19} (1967) 1095}.

\bibitem{Kruskal1}
R.M.~Miura, \emph{{Korteweg‐de Vries Equation and Generalizations. I. A Remarkable Explicit Nonlinear Transformation}}, \href{https://doi.org/10.1063/1.1664700}{\emph{Journal of Mathematical Physics} {\bfseries 9} (1968) 1202}.

\bibitem{Kruskal2}
R.M.~Miura, C.S.~Gardner and M.D.~Kruskal, \emph{{Korteweg‐de Vries Equation and Generalizations. II. Existence of Conservation Laws and Constants of Motion}}, \href{https://doi.org/10.1063/1.1664701}{\emph{Journal of Mathematical Physics} {\bfseries 9} (1968) 1204}.

\bibitem{10.1063/1.1664873}
C.H.~Su and C.S.~Gardner, \emph{{Korteweg‐de Vries Equation and Generalizations. III. Derivation of the Korteweg‐de Vries Equation and Burgers Equation}}, \href{https://doi.org/10.1063/1.1664873}{\emph{Journal of Mathematical Physics} {\bfseries 10} (1969) 536}.

\bibitem{10.1063/1.1665772}
C.S.~Gardner, \emph{{Korteweg‐de Vries Equation and Generalizations. IV. The Korteweg‐de Vries Equation as a Hamiltonian System}}, \href{https://doi.org/10.1063/1.1665772}{\emph{Journal of Mathematical Physics} {\bfseries 12} (1971) 1548}.

\bibitem{10.1063/1.1665232}
M.D.~Kruskal, R.M.~Miura, C.S.~Gardner and N.J.~Zabusky, \emph{{Korteweg‐de Vries Equation and Generalizations. V. Uniqueness and Nonexistence of Polynomial Conservation Laws}}, \href{https://doi.org/10.1063/1.1665232}{\emph{Journal of Mathematical Physics} {\bfseries 11} (1970) 952}.

\bibitem{Laxorig}
P.D.~Lax, \emph{Integrals of nonlinear equations of evolution and solitary waves}, \href{https://doi.org/https://doi.org/10.1002/cpa.3160210503}{\emph{Communications on Pure and Applied Mathematics} {\bfseries 21} (1968) 467}.

\bibitem{Liouville1855}
J.~Liouville, \emph{Note sur l'intégration des équations différentielles de la dynamique, présentée au bureau des longitudes le 29 juin 1853.}, {\emph{Journal de Mathématiques Pures et Appliquées} (1855) 137}.

\bibitem{PhysRevD.11.3424}
R.F.~Dashen, B.~Hasslacher and A.~Neveu, \emph{Particle spectrum in model field theories from semiclassical functional integral techniques}, \href{https://doi.org/10.1103/PhysRevD.11.3424}{\emph{Phys. Rev. D} {\bfseries 11} (1975) 3424}.

\bibitem{PhysRevD.11.2088}
S.~Coleman, \emph{Quantum sine-gordon equation as the massive thirring model}, \href{https://doi.org/10.1103/PhysRevD.11.2088}{\emph{Phys. Rev. D} {\bfseries 11} (1975) 2088}.

\bibitem{ZAMOLODCHIKOV1979253}
A.B.~Zamolodchikov and A.B.~Zamolodchikov, \emph{Factorized s-matrices in two dimensions as the exact solutions of certain relativistic quantum field theory models}, \href{https://doi.org/https://doi.org/10.1016/0003-4916(79)90391-9}{\emph{Annals of Physics} {\bfseries 120} (1979) 253}.

\bibitem{Dubovsky:2015zey}
S.~Dubovsky and V.~Gorbenko, \emph{{Towards a Theory of the QCD String}}, \href{https://doi.org/10.1007/JHEP02(2016)022}{\emph{JHEP} {\bfseries 02} (2016) 022} [\href{https://arxiv.org/abs/1511.01908}{{\ttfamily 1511.01908}}].

\bibitem{Copetti:2024rqj}
C.~Copetti, L.~Cordova and S.~Komatsu, \emph{{Noninvertible Symmetries, Anomalies, and Scattering Amplitudes}}, \href{https://doi.org/10.1103/PhysRevLett.133.181601}{\emph{Phys. Rev. Lett.} {\bfseries 133} (2024) 181601} [\href{https://arxiv.org/abs/2403.04835}{{\ttfamily 2403.04835}}].

\bibitem{DHoker:1982wmk}
E.~D'Hoker and R.~Jackiw, \emph{{Liouville Field Theory}}, \href{https://doi.org/10.1103/PhysRevD.26.3517}{\emph{Phys. Rev. D} {\bfseries 26} (1982) 3517}.

\bibitem{Metsaev:1998it}
R.R.~Metsaev and A.A.~Tseytlin, \emph{{Type IIB superstring action in AdS(5) x S**5 background}}, \href{https://doi.org/10.1016/S0550-3213(98)00570-7}{\emph{Nucl. Phys. B} {\bfseries 533} (1998) 109} [\href{https://arxiv.org/abs/hep-th/9805028}{{\ttfamily hep-th/9805028}}].

\bibitem{PhysRevD.69.046002}
I.~Bena, J.~Polchinski and R.~Roiban, \emph{{Hidden symmetries of the ${\mathrm{AdS}}_{5}\ifmmode\times\else\texttimes\fi{}{\mathrm{S}}^{5}$ superstring}}, \href{https://doi.org/10.1103/PhysRevD.69.046002}{\emph{Phys. Rev. D} {\bfseries 69} (2004) 046002}.

\bibitem{Beisert_2011}
N.~Beisert, C.~Ahn, L.F.~Alday, Z.~Bajnok, J.M.~Drummond, L.~Freyhult et~al., \emph{Review of ads/cft integrability: An overview}, \href{https://doi.org/10.1007/s11005-011-0529-2}{\emph{Letters in Mathematical Physics} {\bfseries 99} (2011) 3–32}.

\bibitem{Ablowitz1993}
M.J.~Ablowitz, S.~Chakravarty and L.A.~Takhtajan, \emph{A self-dual yang-mills hierarchy and its reductions to integrable systems in 1+1 and 2+1 dimensions}, \href{https://doi.org/10.1007/BF02108076}{\emph{Communications in Mathematical Physics} {\bfseries 158} (1993) 289}.

\bibitem{Costello_2018}
K.~Costello, E.~Witten and M.~Yamazaki, \emph{Gauge theory and integrability, i}, \href{https://doi.org/10.4310/iccm.2018.v6.n1.a6}{\emph{Notices of the International Congress of Chinese Mathematicians} {\bfseries 6} (2018) 46–119}.

\bibitem{Costello:2018gyb}
K.~Costello, E.~Witten and M.~Yamazaki, \emph{{Gauge Theory and Integrability, II}}, \href{https://doi.org/10.4310/ICCM.2018.v6.n1.a7}{\emph{ICCM Not.} {\bfseries 06} (2018) 120} [\href{https://arxiv.org/abs/1802.01579}{{\ttfamily 1802.01579}}].

\bibitem{Costello:2019tri}
K.~Costello and M.~Yamazaki, \emph{{Gauge Theory And Integrability, III}},  \href{https://arxiv.org/abs/1908.02289}{{\ttfamily 1908.02289}}.

\bibitem{Gutperle:2014aja}
M.~Gutperle and Y.~Li, \emph{{Higher Spin Lifshitz Theory and Integrable Systems}}, \href{https://doi.org/10.1103/PhysRevD.91.046012}{\emph{Phys. Rev. D} {\bfseries 91} (2015) 046012} [\href{https://arxiv.org/abs/1412.7085}{{\ttfamily 1412.7085}}].

\bibitem{Beccaria:2015iwa}
M.~Beccaria, M.~Gutperle, Y.~Li and G.~Macorini, \emph{{Higher spin Lifshitz theories and the Korteweg-de Vries hierarchy}}, \href{https://doi.org/10.1103/PhysRevD.92.085005}{\emph{Phys. Rev. D} {\bfseries 92} (2015) 085005} [\href{https://arxiv.org/abs/1504.06555}{{\ttfamily 1504.06555}}].

\bibitem{Sasaki:1987mm}
R.~Sasaki and I.~Yamanaka, \emph{{Virasoro Algebra, Vertex Operators, Quantum {Sine-Gordon} and Solvable Quantum Field Theories}}, {\emph{Adv. Stud. Pure Math.} {\bfseries 16} (1988) 271}.

\bibitem{Eguchi:1989hs}
T.~Eguchi and S.-K.~Yang, \emph{{Deformations of Conformal Field Theories and Soliton Equations}}, \href{https://doi.org/10.1016/0370-2693(89)91463-9}{\emph{Phys. Lett. B} {\bfseries 224} (1989) 373}.

\bibitem{ZAMOLODCHIKOV1989641}
A.~Zamolodchikov, \emph{Integrable field theory from conformal field theory},  in \emph{Integrable Systems in Quantum Field Theory and Statistical Mechanics}, M.~Jimbo, T.~Miwa and A.~Tsuchiya, eds., (San Diego), pp.~641--674, Academic Press (1989), \href{https://doi.org/10.1016/B978-0-12-385342-4.50022-6}{DOI}.

\bibitem{Bazhanov:1994ft}
V.V.~Bazhanov, S.L.~Lukyanov and A.B.~Zamolodchikov, \emph{{Integrable structure of conformal field theory, quantum KdV theory and thermodynamic Bethe ansatz}}, \href{https://doi.org/10.1007/BF02101898}{\emph{Commun. Math. Phys.} {\bfseries 177} (1996) 381} [\href{https://arxiv.org/abs/hep-th/9412229}{{\ttfamily hep-th/9412229}}].

\bibitem{Smirnov:2016lqw}
F.A.~Smirnov and A.B.~Zamolodchikov, \emph{{On space of integrable quantum field theories}}, \href{https://doi.org/10.1016/j.nuclphysb.2016.12.014}{\emph{Nucl. Phys. B} {\bfseries 915} (2017) 363} [\href{https://arxiv.org/abs/1608.05499}{{\ttfamily 1608.05499}}].

\bibitem{SISvinolupov_1992}
S.I.~Svinolupov and V.V.~Sokolov, \emph{Factorization of evolution equations}, \href{https://doi.org/10.1070/RM1992v047n03ABEH000895}{\emph{Russian Mathematical Surveys} {\bfseries 47} (1992) 127}.

\bibitem{Svinolupov1982}
S.I.~Svinolupov and V.V.~Sokolov, \emph{Evolution equations with nontrivial conservative laws}, \href{https://doi.org/10.1007/BF01077866}{\emph{Functional Analysis and Its Applications} {\bfseries 16} (1982) 317}.

\bibitem{Sokolov1985}
V.G.~Drinfel'd, S.I.~Svinolupov and V.V.~Sokolov, \emph{Classification of fifth-order evolution equation having an infinite series of conservation laws}, {\emph{Dokl. AN USSR} {\bfseries A10} (1985) 7}.

\bibitem{Mikhailov1991}
A.V.~Mikhailov, A.B.~Shabat and V.V.~Sokolov, \emph{The symmetry approach to classification of integrable equations},  in \emph{What Is Integrability?}, V.E.~Zakharov, ed., (Berlin, Heidelberg), pp.~115--184, Springer Berlin Heidelberg (1991), \href{https://doi.org/10.1007/978-3-642-88703-1_4}{DOI}.

\bibitem{Mikhailov2009}
A.~Mikhailov and V.~Sokolov, \emph{Symmetries of differential equations and the problem of integrability},  in \emph{Integrability}, A.V.~Mikhailov, ed., (Berlin, Heidelberg), pp.~19--88, Springer Berlin Heidelberg (2009), \href{https://doi.org/10.1007/978-3-540-88111-7_2}{DOI}.

\bibitem{meshkov2013integrable}
A.G.~Meshkov and V.V.~Sokolov, \emph{Integrable evolution equations with constant separant}, {\emph{Ufa Math. J.} {\bfseries 4} (2012) 104}.

\bibitem{Heredero:2019arc}
R.H.~Heredero and V.~Sokolov, \emph{{The symmetry approach to integrability: recent advances}},  \href{https://arxiv.org/abs/1904.01953}{{\ttfamily 1904.01953}}.

\bibitem{Lindwasser:2024qyh}
L.W.~Lindwasser, \emph{{On the space of 2d integrable models}}, \href{https://doi.org/10.1007/JHEP01(2025)138}{\emph{JHEP} {\bfseries 01} (2025) 138} [\href{https://arxiv.org/abs/2409.08266}{{\ttfamily 2409.08266}}].

\bibitem{olver1993applications}
P.~Olver, \emph{Applications of Lie Groups to Differential Equations}, Graduate Texts in Mathematics, Springer New York (1993).

\bibitem{zbMATH02644938}
G.~Tzitz{\'e}ica, \emph{Sur une nouvelle classe de surfaces.}, {\emph{C. R. Acad. Sci., Paris} {\bfseries 144} (1907) 1257}.

\bibitem{Zhiber:1979am}
A.V.~Zhiber and A.B.~Shabat, \emph{{Klein-Gordon Equations with a Nontrivial Group}}, {\emph{Sov. Phys. Dokl.} {\bfseries 24} (1979) 607}.

\bibitem{Mikhailov1981}
A.V.~Mikhailov, M.A.~Olshanetsky and A.M.~Perelomov, \emph{Two-dimensional generalized toda lattice}, \href{https://doi.org/10.1007/BF01209308}{\emph{Communications in Mathematical Physics} {\bfseries 79} (1981) 473}.

\bibitem{Sokolov1984class}
V.~Sokolov and A.~Shabat, \emph{Classification of integrable evolution equations}, {\emph{Soviet Scientific Reviews. Section C} {\bfseries 4} (1984) }.

\bibitem{SANDERS1998410}
J.A.~Sanders and J.P.~Wang, \emph{On the integrability of homogeneous scalar evolution equations}, \href{https://doi.org/https://doi.org/10.1006/jdeq.1998.3452}{\emph{Journal of Differential Equations} {\bfseries 147} (1998) 410}.

\bibitem{Klimcik:2008eq}
C.~Klimcik, \emph{{On integrability of the Yang-Baxter sigma-model}}, \href{https://doi.org/10.1063/1.3116242}{\emph{J. Math. Phys.} {\bfseries 50} (2009) 043508} [\href{https://arxiv.org/abs/0802.3518}{{\ttfamily 0802.3518}}].

\bibitem{Klimcik:2014bta}
C.~Klimcik, \emph{{Integrability of the bi-Yang-Baxter sigma-model}}, \href{https://doi.org/10.1007/s11005-014-0709-y}{\emph{Lett. Math. Phys.} {\bfseries 104} (2014) 1095} [\href{https://arxiv.org/abs/1402.2105}{{\ttfamily 1402.2105}}].

\bibitem{Sfetsos:2013wia}
K.~Sfetsos, \emph{{Integrable interpolations: From exact CFTs to non-Abelian T-duals}}, \href{https://doi.org/10.1016/j.nuclphysb.2014.01.004}{\emph{Nucl. Phys. B} {\bfseries 880} (2014) 225} [\href{https://arxiv.org/abs/1312.4560}{{\ttfamily 1312.4560}}].

\bibitem{Zarembo:2017muf}
K.~Zarembo, \emph{{Integrability in Sigma-Models}},  \href{https://arxiv.org/abs/1712.07725}{{\ttfamily 1712.07725}}.

\bibitem{Orlando:2019his}
D.~Orlando, S.~Reffert, J.-i.~Sakamoto, Y.~Sekiguchi and K.~Yoshida, \emph{{Yang\textendash{}Baxter deformations and generalized supergravity\textemdash{}a short summary}}, \href{https://doi.org/10.1088/1751-8121/abb510}{\emph{J. Phys. A} {\bfseries 53} (2020) 443001} [\href{https://arxiv.org/abs/1912.02553}{{\ttfamily 1912.02553}}].

\bibitem{Klimcik:2021bjy}
C.~Klim\v{c}\'\i{}k, \emph{{Brief lectures on duality, integrability and deformations}}, \href{https://doi.org/10.1142/S0129055X21300041}{\emph{Rev. Math. Phys.} {\bfseries 33} (2021) 2130004} [\href{https://arxiv.org/abs/2101.05230}{{\ttfamily 2101.05230}}].

\bibitem{Hoare:2021dix}
B.~Hoare, \emph{{Integrable deformations of sigma models}}, \href{https://doi.org/10.1088/1751-8121/ac4a1e}{\emph{J. Phys. A} {\bfseries 55} (2022) 093001} [\href{https://arxiv.org/abs/2109.14284}{{\ttfamily 2109.14284}}].

\bibitem{Ferko:2024ali}
C.~Ferko and L.~Smith, \emph{{Infinite Family of Integrable Sigma Models Using Auxiliary Fields}}, \href{https://doi.org/10.1103/PhysRevLett.133.131602}{\emph{Phys. Rev. Lett.} {\bfseries 133} (2024) 131602} [\href{https://arxiv.org/abs/2405.05899}{{\ttfamily 2405.05899}}].

\bibitem{Bielli:2024fnp}
D.~Bielli, C.~Ferko, L.~Smith and G.~Tartaglino-Mazzucchelli, \emph{{Auxiliary Field Sigma Models and Yang-Baxter Deformations}},  \href{https://arxiv.org/abs/2408.09714}{{\ttfamily 2408.09714}}.

\end{thebibliography}\endgroup
\end{document}